\documentstyle[aps,prb]{revtex} 
\input epsf.sty

\begin{document}
\setcounter{figure}{0} 
\renewcommand{\theequation}{\arabic{section}.\arabic{equation}} 

\twocolumn[ 
\hsize\textwidth\columnwidth\hsize\csname@twocolumnfalse\endcsname 
 
\draft 

\title{Gauge Theory of Composite Fermions:
\\
Particle-Flux Separation in Quantum Hall Systems} 
\author{Ikuo Ichinose\cite{ichinose}} 
\address{Department of Electrical and Computer Engineering,
Nagoya Institute of Technology, 
Nagoya, 466-8555 Japan 
} 
\author{Tetsuo Matsui\cite{matsui}} 
\address{Department of Physics, Kinki University, 
Higashi-Osaka, 577-8502 Japan 
} 
\date{\today}  
 
\maketitle

\begin{abstract}   
Fractionalization phenomenon of electrons in quantum Hall states is 
studied in terms of U(1) gauge theory. We focus on the Chern-Simons(CS) 
fermion description of the quantum Hall effect(QHE) at the filling factor 
$\nu=p/(2pq\pm 1)$, and show that the successful composite-fermions(CF) 
theory of Jain acquires a solid theoretical basis, which  we call
particle-flux separation(PFS). PFS can be studied efficiently by a  
gauge theory and characterized as a deconfinement phenomenon in the 
corresponding gauge dynamics. The PFS takes place at low temperatures, 
$T \leq T_{\rm PFS}$, where each electron or CS fermion splinters off 
into two quasiparticles, 
a fermionic chargeon and a bosonic fluxon. The chargeon is nothing but 
Jain's CF, and the fluxon carries $2q$ units of CS fluxes. 
At sufficiently low temperatures $T \leq T_{\rm BC} ( < T_{\rm PFS})$,
fluxons Bose-condense uniformly and (partly) cancel the external magnetic 
field, producing the correlation holes. This partial cancellation validates 
the mean-field theory in Jain's CF approach. FQHE takes place at 
$T < T_{\rm BC}$ as a joint effect of (i) integer QHE of chargeons under 
the residual field $\Delta B$ and (ii) Bose condensation of fluxons.
We calculate the phase-transition temperature $T_{\rm PFS}$ and the CF mass.
Repulsive interactions between electrons are essential to establish PFS. 
PFS is a counterpart of the charge-spin separation in the t-J model of 
high-$T_{\rm c}$ cuprates in which each electron dissociates into holon 
and spinon. Quasiexcitations and resistivity in the PFS state are also 
studied. The resistivity is just the sum of contributions of chargeons 
and fluxons, and $\rho_{xx}$ changes
its behavior at $T = T_{\rm PFS}$, reflecting the change of
quasiparticles from chargeons and fluxons at $T < T_{\rm PFS}$ to 
electrons at $T_{\rm PFS} < T$. 
\end{abstract} 
 

]
 
\setcounter{footnote}{0}
\setcounter{equation}{0} 
\section{Introduction}
Fractional quantum Hall effect(FQHE) is one of the most interesting 
phenomena in the current condensed matter physics.
As a result of Coulomb interactions between electrons, several excentric 
quasiparticles may appear as low-energy excitation modes.
Among them,
composite fermions (CFs) proposed by Jain \cite{Jain} 
play a very important role for a unified view of the FQHE
at the states with the electron filling factor $\nu=p/(2pq\pm 1)$
($p, q$ are positive integers).
Experiments support the CF picture not only in these FQH states,
but also in the compressible states like $\nu=1/2$.
In theoretical studies of CF, the Chern-Simons(CS) gauge
theory is often employed.
However, there are no transparent and consistent theoretical
explanations why the CF picture works so nicely.
In particular, in the CS gauge theory of CF given so far, it is not clear 
how to treat the CS constraints consistently, although
this problem is essential to determine the low-energy quasiparticles
and discuss the (in)stability of CFs.
Another problem is to calculate the CF mass, which is to be determined 
by the Coulomb interactions between electrons.
{\em Nonperturbative} study on the CS gauge theory is necessary for
solving these problems, though most of the studies use  perturbative 
calculations with respect to the CS gauge field assuming
a small gauge coupling.\cite{LF,HLR}  

Recently, some papers on the low-energy properties of the QHE and/or
the CS gauge theory appeared.
Shankar and Murthy (SM)\cite{SM} tried to separate intra-Landau
level (LL) excitations from inter-LL excitations by enlarging the Hilbert
space and redefining field variables.
However, as a result of these redefinitions, the resultant field operators
satisfy very complicated nonlocal commutation relations, though SM just
ignore these complexities.
Therefore these field operators cannot be capable to describe 
quasiexcitations at low energies.
In other words,
enlargement of the Hilbert space and redefinition of the field operators
are not sufficient to describe low-energy physics of FQHE and its closely
related states.

In the SM approach, there appears a local gauge symmetry and they impose
a constraint on physical states, which is similar to the Gauss' law 
constraint in quantum electrodynamics(QED).
However, the way  how this gauge symmetry is realized
dynamically (e.g., in confienement phase or in deconfinement phase) 
is crucial for the physical properties of the system at
low energies.
For example, in QED, the field operators of electrons and photons
in the {\em Heisenberg picture} satisfy the Gauss' law, 
but the {\em asymptotic fields}
describing incoming and outgoing particles {\em do not}.
This stems from the fact that the
gauge symmetry in QED is realized in the deconfinement-Coulomb phase.
As a result, electrons and photons appear as quasi-free particles and
their interactions can be treated by the perturbative calculations
with respect to the {\it small} gauge-coupling constant at low energies.

In Jain's idea of CF,\cite{Jain} notion and dynamics of 
fictitious (i.e., CS) magnetic fluxes 
are very important to understand the FQHE and $\nu = 1/2$ states
intuitively.
In the previous papers\cite{PFS1,PFS2}, 
we studied an electron system 
in a strong magnetic field at $\nu = 1/2$.
In Ref.\cite{PFS1} we made  the usual CS gauge transformation
to electrons and studied the resulting CS gauge theory 
of self-interacting fermions. We point out the
possibility of separation phenomenon between particle and flux
degrees of freedom, which we called particle-flux separation (PFS).
In Ref.\cite{PFS2}, we associated a particle picture
to flux degrees of freedom by writing
the CS fermion operator as a product of two operators of 
new particles named chargeon and fluxon by enlarging the Hilbert space 
and imposing a local constraint.
The chargeons are just the fermions that describe Jain's CFs,
exhibitting IQHE at low temperatures,
whereas the fluxons are bosons that describe CS flux degrees of
freedom and associated correlated holes.
We showed that PFS takes place at low 
temperatures ($T \leq T_{\rm PFS}$) where a CS fermion fractionalizes 
into a chargeon and a fluxon, and these ``constituents" behave 
as quasi-free low-energy excitations there.

PFS is closely related with the 
charge-spin separation (CSS) in high-$T_{\rm c}$ cuprates, in which each 
electron dissociates into holon and spinon.\cite{CSS,IMS,IMO}
Both the PFS and CSS are understood as  deconfinement phenomena
of dynamical gauge fields that appear as a result of introducing 
constituents of an electron, the chargeon
and fluxon in PFS and the  holon and spinon in CSS.

\begin{figure} 
  \begin{picture}(180,270) 
    \put(0,0){\epsfxsize 180pt \epsfbox{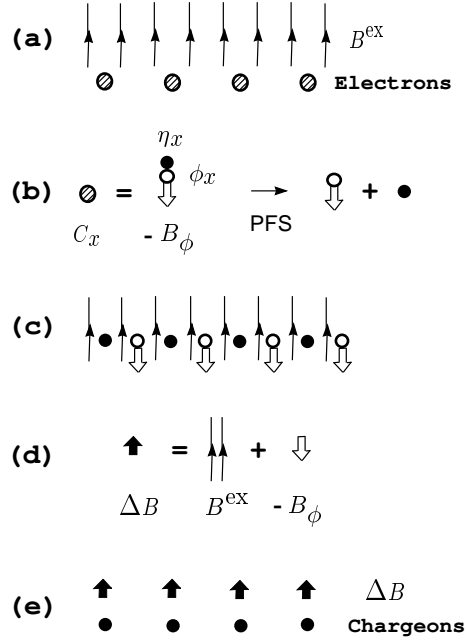}} 
  \end{picture} 
\caption{
  Illustration of chargeon-fluxon theory of CF and PFS.
(a) Electrons in 
a magnetic field $B^{\rm ex}$. Thin arrows are $B^{\rm ex}$
and hatched circles are electrons $C_x$.
 (b) As described in (\ref{CStrans}), each electron is regarded
as a composite of CS fluxes $-B_\phi$ (a thick arrow)
expressed by the phase factor in (\ref{CStrans})
 and  a CS fermion $\psi_x$.  
 Then, as shown in (\ref{CFoperator}), 
 $\psi_x$ is viewed
 as a composite of a fluxon $\phi_x$ (an open circle)
and a chargeon $\eta_x$ (a filled circle).
Each fluxon is accompanied with CS fluxes.  
When the  PFS takes place at $T < T_{\rm PFS}$, chargeons and 
fluxons with CS fluxes dissociate and behave independently. 
(c) In PFS states, 
chrgeons feel a reduced magnetic field (d) $\Delta B$ (a thick
black arrow) given by  (\ref{Bphi})
and (\ref{deltab}). (e) At $T < T_{\rm BC} (< T_{\rm PFS})$
fluxons Bose-condense and the system reduces to fermionic
chargeons in a constant magnetic field $\Delta B$.
At $T < T_{\rm BC} $, FQHE takes place at $\nu = p/(2pq \pm 1)$
as a joint effect of (I) IQFE of chargeons in $\Delta B$ and
(II) Bose condensation of fluxons.
}
\label{pfs1}
\end{figure}

In our recent letter \cite{PFS3} we generalized the gauge theory of PFS 
at $\nu = 1/2$ of Ref.\cite{PFS1,PFS2} to the cases of $\nu = p/(2pq \pm1)$
($p,q:$ positive integers). The case of $\nu = 1/2$ 
is viewed as the limit $p \rightarrow \infty$ at $q=1$.
In the PFS states, bosonic fluxons may  Bose condense at $T \leq 
T_{\rm BC} (< T_{\rm PFS})$. The resulting uniform CS magnetic field
(partly) cancels the external magnetic field, and chargeons move
in this reduced field. This partial cancellation of magnetic field
just validates Jain's idea that FQHE is nothing but the integer QHE
of CFs, where chargeons are nothing but CFs.
In Fig.\ref{pfs1} we illustrate the basic idea of our chargeon-fluxon 
approach to FQHE.

In this paper, we shall present detailed and self-contained
 account of the gauge theory 
of composite fermions summerized in our letter.\cite{PFS3}
We extend also the previous calculations of the 
transition temperature 
$T_{\rm PFS}$ and masses of chargeon and fluxon at $\nu = 1/2$ 
to the general fillings.
Furthermore, we investigate the physical 
properties of the low-energy excitations in 
PFS states in the present picture. 

The paper is organized as follows;
In Sec.II, we introduce our model.
 Field operators of chargeons and fluxons are introduced to describe 
strongly-correlated electron systems in a magnetic field.
Chargeons and fluxons are quantized as ordinary fermions and bosons,
respectively.
A local gauge symmetry emerges as 
a result of the chargeon-fluxon representation of electron, and
a gauge field is introduced as an auxiliary field, which mediates
the interaction between chargeon and fluxon in an electron. 
In Sec.III, an effective action or a Ginzburg-Landau(GL) theory
of the dynamical gauge field is obtained by integrating out chargeon
and fluxon variables by the hopping expansion.
By referring to the established knowledge of lattice gauge theory,
it is concluded that a confinement-deconfinement phase transition of
the gauge dynamics takes place at the transition temperature
$T_{\rm PFS} ( > 0)$, and the PFS (deconfinement phase) is
realized {\it below}  $T_{\rm PFS}$.
In Sec.IV, the ground states, 
quasiexcitations, and the EM transport properties  in 
PFS states are discussed.
The ground state is a direct product of the chargeon ground state
and Bose condensate of fluxons, which is shown to be just 
the Laughlin states for $p=1$.
As a result of the local gauge invariance, we obtain
a formula of resistivity similar to the Ioffe-Larkin formula 
in the holon-spinon theory of high-$T_{\rm c}$ cuprates; the resistivity
tensor is just the summation of chargeon contribution and fluxon contribution.This formula gives rise to the results that agree with the experimantal
observations. At $T = T_{\rm PFS}$, $\rho_{xx}$ changes
its behavior   reflecting the change of
quasiparticles from chargeons and fluxons at $T < T_{\rm PFS}$ to 
electrons at $T_{\rm PFS} < T$. 

Sec.V is devoted to discussion. We explain
flaws of SM theory in some details, e.g., insufficient
treatment of noncommuting operaotrs. These flaws are 
removed in our theory, which is characterized as
a detailed study of  the dynamics of the relevant 
degrees of freedom in a framework of second-quantized field theory.

\section{Model}
\setcounter{equation}{0}

\subsection{Electrons on a lattice}

Let us consider a two-dimensional system of electrons
in a parpendicular constant magnetic field $B^{\rm ex}$.
Instead of working in the continuum space, we put 
the system on a two-dimensional square lattice as a 
way of regularization, which
is useful for nonperturbative study of gauge dynamics
as demonstrated in lattice gauge theory by Wilson.\cite{wilson}
The lattice model below is regarded as an effective system of
renormalization-group theory, and 
the main results obtained below, e.g., the existence of PFS, 
are not the artifacts of introduction of a lattice but survive 
in the continuum. For example, the critical temperature 
$T_{\rm PFS}$ is a physical quantity that is renormalization-group 
invariant.(We discuss more on this point at the end of Sec.III.) 
For definiteness, in numerical estimation, we take the lattice spacing 
$a$ as $a\sim \ell$ where 
\begin{eqnarray}
\ell =\frac{1}{\sqrt{e B^{\rm ex}}}
\end{eqnarray}
is the magnetic length. In this paper we use the 
units $\hbar =1$ and $c=1$. 

We start with
the electron annihilation operator $C_x$
on the site $x$ (it has a definite spin component due to the Zeeman
effect), satisfying the canonical anticommutaiton relations,
\begin{eqnarray}
&&[C_x, C_y^\dagger]_+ =\delta_{xy}, \
[C_x, C_y]_+ =0.
\label{algebraC}
\end{eqnarray}
The Hamiltonian is written as
\begin{eqnarray}
H_C &=&-{1\over 2ma^2}\sum_{x}\sum_{j=1}^2\Big(C^\dagger_{x+j}
e^{-ie(A_{xj}^{\rm ex}+a_{xj})}
C_x+{\rm H.c.}\Big)\nonumber\\
&& +H_{\rm int}(C_x^\dagger C_x),
\label{HC}
\end{eqnarray}
where $m$ is the effective electron mass (i.e., the band mass).
The first term describes the process of electron hopping from 
the site $x$ to its nearest-neighbor $x+j$ (we
use the direction index $j=1,2$ also as the lattice unit vector
in the $j$-th direction) under the EM vector potential
$A^{\rm ex}_{xj}$  
for $B^{\rm ex}a^2=\epsilon_{ij}\nabla_i A^{\rm ex}_{xj}$ 
($\epsilon_{12} =1$ and 
$\nabla_i F(x) \equiv F(x+i)-F(x)$).
We included also the dynamical (fluctuating) EM vector potential, 
$a_{xj}$, in order to derive the formula of the EM current later.
By  making the field rescalings,
\begin{eqnarray}
A^{\rm ex}_{xj} \rightarrow a A_j(x),\; a_{xj} \rightarrow a a_j(x),\; 
C_{x} \rightarrow a C(x),
\end{eqnarray}
and taking a naive
continuum limit, $a \rightarrow 0$, 
the first term reduces to a well known continuum Hamiltonian
$(2m)^{-1}  \int d^2x |D_j C(x)|^2,\ D_j\equiv \partial_j 
-ieA^{\rm ex}_j(x)-ie a_j(x)$ (up to a chemical potential term).
The second term $H_{\rm int}$ represents
repulsive interactions between electrons. 
Its  explicit form is specified later.

Note that $A^{\rm ex}_{xj}$ is defined on the link $(x,x+j)$ and
its exponentiated phase factor $U^{\rm ex}_{xj} \equiv \exp(-ie
A^{\rm ex}_{xj})$ is a U(1) link variable of lattice gauge 
theory.\cite{wilson}
In terms of $U_{xj}$ the local (i.e., site-dependent) 
U(1) EM gauge transformation is written as 
\begin{eqnarray}
C_x &\rightarrow& G^{\rm EM}_x C_x,\nonumber\\
U_{xj}e^{-iea_{xj}}&\rightarrow& G^{\rm EM}_{x+j}U_{xj}e^{-iea_{xj}}
{G^{\rm EM}_x}^\dagger,\ G_x \in U(1). 
\label{EMgauge}
\end{eqnarray}

\subsection{CS fermions}

Let us express $C_x$ in terms of CS fermion operator $\psi_x$ 
as follows;
\begin{equation}
C_x=\exp[2iq \sum_y \theta_{xy} \psi^\dagger_y\psi_y]\psi_x,
\label{CStrans}
\end{equation}
where $\theta_{xy} [\equiv \tan^{-1}((x_2-y_2)/(x_1-y_1))]$ 
is the multi-valued azimuthal angle of the vector
$\vec{x}-\vec{y}$
on the lattice.\cite{fradkin} 
From (\ref{CStrans}) $\psi_x$ satisfy  
\begin{eqnarray}
&&[\psi_x, \psi_y^\dagger]_+ =\delta_{xy}, \
[\psi_x, \psi_y]_+ =0.
\label{algebrapsi}
\end{eqnarray}
Eq.(\ref{CStrans}) indicates that each electron is viewed as a 
composite of $2q$ flux quanta ($q=1,2,3,\cdots$) and a 
CS fermion.\cite{fradkin} 
 
Then the Hamiltonian (\ref{HC}) is rewritten as
\begin{eqnarray}
H_\psi &=&-{1\over 2ma^2}\sum_{x,j}\Big(\psi^\dagger_{x+j}
e^{i(A_{xj}^{\rm CS}-eA_{xj}^{\rm ex}-ea_{xj})}
\psi_x+{\rm H.c.}\Big)\nonumber\\
&& +H_{\rm int}(\psi_x^\dagger\psi_x),
\label{Hpsi}
\end{eqnarray}
where the CS gauge field is introduced as 
\begin{eqnarray}
A_{xj}^{\rm CS}=2q\epsilon_{ij}
\sum_y \nabla_i\theta_{xy} \cdot \psi_y^\dagger\psi_y.
\label{ACS}
\end{eqnarray}
Note $H_{\rm int}(C^\dagger_x C_x) = H_{\rm int}(\psi^\dagger_x \psi_x)$ 
due to the relation between the number operators, $C^\dagger_x C_x =
\psi^\dagger_x \psi_x$.
The filling factor $\nu$ is defined as usual by 
\begin{eqnarray}
\nu &=&\frac{2\pi\rho}{eB^{\rm ex}},\ \rho= \frac{n}{a^2},\nonumber\\
n& \equiv& \langle C^\dagger_xC_x\rangle 
=\langle \psi^\dagger_x\psi_x\rangle,
\label{nu}
\end{eqnarray}
where $n$ is the average electron number {\it per site}.
Eqs.(\ref{Hpsi}) and (\ref{ACS}) show that
the CS fermions $\psi_x$ not only
minimally couple with the CS gauge field but also they themselves
are the sources of $A^{\rm CS}_{xj}$, hence producing
the CS magnetic fluxes. In fact, eq(\ref{ACS}) leads to the relation,  
\begin{eqnarray}
B^{\rm CS}_{x} &\equiv & \epsilon_{ij}\nabla_i A^{\rm CS}_{xj} = 
4\pi q \psi^\dagger_x \psi_x,
\label{CSconstraint}
\end{eqnarray}
so each CS fermion accompanies with 2q units of CS flux.

One may conceive that Jain's CF idea can be
realized if the phase factor   and the CS fermion $\psi_x$
in (\ref{CStrans})``sepatate" dynamically. 
However, since both of these quantities
are desctibed by the same variables $\psi_x$, such a separation
is not straightforward. We need to prepare a set of independent variables
to describe the phase (CS fluxes) and the fermion.
This leads us to introduce the chargeon and fluxon  
variables, which is the subject of the next subsection.

\subsection{Chargeons and fluxons}

Let us rewrite the CS fermion operator $\psi_x$ as a composite field of  
a canonical {\em fermion} operator $\eta_x$ and a canonical 
{\em boson} operator $\phi_x$,
\begin{eqnarray}
\psi_x=\phi_x\eta_x.
\label{CFoperator}
\end{eqnarray}
We call particles described by $\eta_x$  and 
$\phi_x$ chargeons and fluxons, respectively.
Meaning of this terminology becomes clear shortly.
They satisfy
\begin{eqnarray}
&&[\phi_x, \phi_y^\dagger] =\delta_{xy}, \ 
[\phi_x, \phi_y] =0, \nonumber\\
&&[\eta_x, \eta_y^\dagger]_+ =\delta_{xy}, \
[\eta_x, \eta_y]_+ =0, \nonumber\\
&&[\phi_x, \eta_y] = [\phi_x, \eta_y^\dagger] =0.
\label{algebra2}
\end{eqnarray}
To keep the equivalence of $C_x$ representation and $\phi_x-\eta_x$
 representation, we impose the following local constraint on 
the physical states $|{\rm Phys}\rangle$;
\begin{equation}
\eta^\dagger_x\eta_x|{\rm Phys}\rangle=\phi^\dagger_x\phi_x|
{\rm Phys}\rangle \ \ {\rm for}\ {\rm each}\ x.
\label{const}
\end{equation}
Then there is the following one-to-one correpondence between the
states made of $\psi_x$ and the physical states made of $\phi_x$ and
$\eta_x$;
\begin{eqnarray}
\psi_x|0\rangle_\psi &=&0,\   \eta_x|V\rangle_\eta =
\phi_x|V\rangle_\phi =0, \nonumber\\
|0\rangle _\psi &=& |V\rangle\equiv |V\rangle_\eta|V\rangle_\phi,
 \nonumber\\  
|1\rangle_\psi &\equiv& \psi^\dagger_x|0\rangle_\psi =
\eta_x^\dagger\phi_x^\dagger|V\rangle.
\label{state}
\end{eqnarray}
One may check the product $\phi_x\eta_x$ satisfies the
canonical anticommutation relation  (\ref{algebrapsi}) of $\psi_x$
in the physical subspace defined above.

By substituting (\ref{CFoperator}) to $H_{\psi}$ of (\ref{Hpsi})
 the Hamiltonian is rewritten 
in terms of $\eta_x$ and $\phi_x$  as
\begin{eqnarray}
H_{\eta\phi}&=&-{1\over 2ma^2}\sum_{x,j}
\Big(\eta_{x+j}^\dagger\phi_{x+j}^\dagger
W_{x+j}M_{x+j}M^\dagger_xW^\dagger_xe^{-iea_{xj}}\phi_x\eta_x \nonumber\\
&&+{\rm H.c.}\Big)+H_{\rm int}(\eta_x^\dagger\eta_x
\phi_x^\dagger\phi_x) 
-\sum_x(\mu_\eta\eta_x^\dagger\eta_x+\mu_\phi\phi_x^\dagger\phi_x)
\nonumber \\
&&-\sum_x\lambda_x(\eta_x^\dagger\eta_x-\phi_x^\dagger\phi_x),
\label{Hetaphi}
\end{eqnarray}
where $\lambda_x$ is the Lagrange multiplier to enforce
the constraint (\ref{const}). $W_x$ and $M_x$ are U(1)
phase factors defined as  
\begin{eqnarray}
W_x&=&\exp \Big[2iq\sum_y\theta_{xy}\Big(\phi_y^\dagger\phi_y-
n\Big)\Big], \label{W} \\
M_x&=&\exp \Big[i \sum_y\theta_{xy}(2q-{1\over \nu})n\Big],
\label{M}
\end{eqnarray}
where we have replaced $\psi^\dagger_y \psi_y$ in $A^{\rm CS}_{xj}$
by $\phi^\dagger_y\phi_y$ using the relations 
$\psi^\dagger_x\psi_x=\phi^\dagger_x\phi_x =\eta^\dagger_x\eta_x$
that hold due to the constraints.
The chemical potentials $\mu_\eta$ and $\mu_\phi$ are introduced
to enforce the conditions $\langle \eta^\dagger_x\eta_x \rangle = 
\langle \phi^\dagger_x\phi_x \rangle = n$.

The original electron operator $C_x$ is expressed 
in terms of $\eta_x$ and $\phi_x$ as
\begin{equation}
C_x=\exp [2iq\sum_y\theta_{xy}\phi^\dagger_y\phi_y]\phi_x\eta_x.
\label{electronop}
\end{equation}
Here we have expressed the CS phase factor in terms of $\phi_x$, i.e.,
we have assigned so that each  $\phi_x$ carries CS $2q$ flux quanta,
while $\eta_x$ does not. Thus the fluxons accompany a CS field,
the average value $B_\phi$ of which 
is given by using (\ref{CSconstraint}) as
\begin{eqnarray}
B_\phi &=& \frac{\langle B_x^{\rm CS} \rangle}{e}
 = \frac{4\pi q n }{e a^2},
\label{Bphi}
\end{eqnarray} 
where we divided $B_x^{\rm CS}$ by $e$ so that $B_\phi$ has the dimension
of a magnetic field.

By using the constraint (\ref{const}), two expressions of the electron
operators, (\ref{CStrans}) and (\ref{electronop}), are equivalent.
Eq.(\ref{electronop}) shows that an electron is viewed as a composite of
a chargeon $\eta_x$ and a fluxon 
$\phi_x$ that bears $2q$-flux quanta.
In other words, the {\em chargeon} $\eta_x$ is a 
{\em composite} of an electron
and $2q$-flux quanta in the  direction opposite to $B^{\rm ex}$.
This implies that the chargeon in the present formalism
may be regarded as Jain's CF.
However, we stress that, in order to  justify  the CF picture 
as a physically correct picture,
the PFS must take place dynamically, as we explained in 
Sec.I.

We should also remark here that
the CS fermion and the electron operators are invariant under the following
``gauge transformation" of the chargeon and fluxon;
\begin{equation}
(\eta_x,\phi_x) \rightarrow (e^{i\alpha_x}\eta_x, e^{-i\alpha_x}\phi_x),
\label{gaugetr}
\end{equation}
where $\alpha_x$ is an arbitrary function of $x$.
 As we shall see later, this local gauge symmetry plays a crucial role
for the PFS transition and for the EM transport properties of the present
system.

\subsection{Gauge theory of chargeons and fluxons}
\label{gtcf}

We are mainly interested in the structure of
the ground state and the low-energy excitations of the
system $H_{\eta\phi}$ of (\ref{Hetaphi}).
For this purpose, we employ the Lagrangian path-integral formalism, 
since this formalism is suitable to introduce 
a gauge field as an auxiliary field and to obtain an effective
gauge theory. By using this gauge theory and examining
its gauge dynamics, one can study
 the possibility of PFS in a natural and convincing manner.

The partition funciton $Z$ at the temperature $T$ is expressed
by a path integral \cite{sakita} as
\begin{eqnarray}
Z&\equiv& {\rm Tr} \exp(-\beta H_{\phi\eta}) = 
\int [d\eta] [d\phi][d\lambda]\ \exp\Big(\int^\beta_0d\tau L(\tau)\Big),
\nonumber\\
L&=& -\sum_x \eta_x^\dagger(\partial_\tau+i\lambda_x-\mu_\eta)\eta_x
\nonumber\\
&&-\sum_x \phi_x^\dagger(\partial_\tau-i\lambda_x-\mu_\phi)\phi_x  
\nonumber\\
&&+{1\over 2m}\sum_{x,j}
\Big(\eta_{x+j}^\dagger\phi_{x+j}^\dagger
W_{x+j}M_{x+j}M^\dagger_xW^\dagger_xe^{-iea_{xj}}\phi_x\eta_x \nonumber\\
&&+{\rm H.c.}\Big)-H_{\rm int}(\eta_x^\dagger\eta_x
\phi_x^\dagger\phi_x), \nonumber\\
&&\hspace{-0.5cm}
[d\eta][d\phi][d\lambda] \equiv \prod_{\tau} \prod_x d\eta_x(\tau) 
d\phi_x(\tau) d\lambda_x(\tau),
\label{Z1}
\end{eqnarray}
where $\tau\in [0,\beta (\equiv (k_{\rm B}T)^{-1})]$ 
is the imaginary time. The  fields in $L(\tau)$ are
functions of $\tau$; $\eta_x(\tau), \phi_x(\tau),
\lambda_x(\tau)$.
$\eta_x(\tau)$ is a Grassmann number, 
$\phi_x(\tau)$ is a complex number, and $\lambda_x(\tau)$ is a real number.

We decouple the third line of $L$ by introducing
a complex link auxiliary field\cite{sakita} $V_{xj}$ 
on the link $(x,x+j)$ by using the formula for arbitrary $J_{xj}$, 
\begin{eqnarray}
&&\int dV_{xj} \exp\Big(\Delta \tau \Big[ -{|V_{xj}|^2 \over 2ma^2} + 
(V^\dagger_{xj} J_{xj} + {\rm H.c.}) \Big]
\Big) \nonumber\\ 
&\propto& \exp\Big(\Delta \tau\ 2ma^2|J_{xj}|^2\Big).
\label{auxiliary}
\end{eqnarray}
Then we  have
\begin{eqnarray}
Z&=& \int [d\eta][d\phi][d\lambda][dV]\ 
\exp\Big(\int^\beta_0d\tau L_{V}(\tau)\Big),
\nonumber\\
L_V&=&-\sum_x \eta_x^\dagger(\partial_\tau+i\lambda_x-\mu_\eta)\eta_x
\nonumber\\
&-& \sum_x \phi_x^\dagger(\partial_\tau-i\lambda_x-\mu_\phi)\phi_x  
+\sum \Big(V^\dagger_{xj} J_{xj} + {\rm H.c.}\Big)\nonumber\\
&-&{1\over 2ma^2}\sum_{x,j}|V_{xj}|^2
- H_4 -H_{\rm int}(\eta_x^\dagger\eta_x
\phi_x^\dagger\phi_x), \nonumber\\
H_{4} &\equiv& \sum_{x,j }\Big({\gamma^2\over 2m}
 \phi_{x+j}^\dagger\phi_{x+j}\phi_x^\dagger\phi_x
-{1\over 2ma^2\gamma^2} \eta_{x+j}^\dagger\eta_{x+j}\eta_x^\dagger\eta_x
\Big),  \nonumber \\
J_{xj}&\equiv& {1 \over 2ma^2}\Big(\gamma\phi_{x+j}W_{x}e^{ieca_{xj}}
W^\dagger_{x+j}
\phi_x^\dagger  \nonumber \\
&& \; \; +{1\over \gamma}\eta_{x+j}^\dagger M_{x+j} 
e^{-ie(1-c) a_{xj}}
M^\dagger_x\eta_x\Big).
\label{J}
\end{eqnarray} 
Here $\gamma$ is a real
parameter, which measures the ratio of the chargeon and fluxon 
masses as we shall see in (\ref{effectivemass}). 
We determine its value later in
(\ref{gamma})  of Sec.\ref{repulsiveint}.
$\gamma$ 
$c$ is an {\em arbitrary} real constant that appears in the EM charges
$Q_\phi$ of $\phi_x$ and $Q_\eta$ of $\eta_x$, which are read off
from the couplings to $a_{xj}$ in $L_V$ as   
\begin{equation}
Q_\phi=ce,\; Q_\eta=(1-c)e.
\label{EMcharge}
\end{equation}
In Sec.4C, we shall discuss some consequences of this arbitrariness
in the EM transport properties; we shall see that the physical
results are independent of the value of $c$.

To see the physical meaning of the auxiliary field
$V_{xj}$, we use the following trivial identity; 
\begin{eqnarray}
&&\int d^2V_{xj} \frac{\partial}{\partial V^\dagger_{xj}}
\exp\Big(\Delta \tau \Big[ -{|V_{xj}|^2 \over 2ma^2} + 
(V^\dagger_{xj} J_{xj} + {\rm H.c.}) \Big]\Big) \nonumber\\
&=&\int d^2V_{xj}\Delta \tau  \Big[ 
-\frac{V_{xj}}{2ma^2} +J_{xj}
\Big]\nonumber\\
&&\times
\exp\Big(\Delta \tau \Big[ -{|V_{xj}|^2 \over 2ma^2} + 
(V^\dagger_{xj} J_{xj} + {\rm H.c.}) \Big]\Big)\nonumber\\
&=&0,
\end{eqnarray}
(the surface terms vanish) 
to obtain the relation (equations of motion),
\begin{eqnarray}
\langle V_{xj} \rangle &=& 2ma^2 \langle J_{xj} \rangle,
\end{eqnarray}
which shows that $V_{xj}$ describes the hopping amplitudes
of $\eta_x$ and $\phi_x$. 
We note that the way of decoupling
in (\ref{J}) is slightly different from our previous papers.\cite{PFS2} 
The present expression is more suitable to discuss
PFS since the current $J_{xj}$ coupled to $V_{xj}$
is just the sum of the $\eta_x$ part and the $\phi_x$ part
without mixing terms.

From (\ref{J}), 
$A_{x0} \equiv \lambda_x$ 
and $A_{xj}$ defined through $V_{xj} \equiv |V_{xj}| U_{xj},\ 
U_{xj} \equiv \exp(iA_{xj})$ can be 
regarded as the time and transverse component of a gauge field 
$A_{x\mu}$ ($\mu = 0(\tau), 1,2$), respectively.
Actually, under the U(1) local gauge transformation (\ref{gaugetr})
with $\tau$-dependent $\alpha_x(\tau)$,
they transform as
\begin{equation}
\lambda_x\rightarrow \lambda_x-\partial_\tau \alpha_x,\;\;
A_{xj}\rightarrow A_{xj}+\nabla_j\alpha_x.
\label{gaugetr2}
\end{equation}
The Lagrangian $L_V$ of (\ref{J}) is invariant under (\ref{gaugetr2}).
Thus the system has a local gauge invariance, and one may regard
the system as a lattice gauge theory.\cite{wilson}

 A lattice gauge theory  has generally two 
possibilities to realize its gauge dynamics;\\

\noindent (i) Confinement phase:

Here the gauge-field fluctuations, $\Delta A_{xj}$, are large
and random i.e.,
\begin{eqnarray}
\langle (\Delta A_{xj})^2 \rangle = \infty, \ \langle \exp(iA_{xj}) \rangle = 0,
\end{eqnarray} 
and only the gauge-invariant (i.e., charge-neutral) 
objects may appear as physical excitations.\\

\noindent
(ii) Deconfinement phase (like Coulomb phase and Higgs phase):

Here $\Delta A_{xj}$ 
are small and can be treated
in perturbation theory, 
\begin{eqnarray}
\langle (\Delta A_{xj})^2 \rangle \sim 0, \ 
\langle \exp(iA_{xj}) \rangle \sim 1,
\end{eqnarray}
and the {\it gauge-variant} objects may appear as excitations.
(For more detailed discussion on this point, see Sec.VI of the
first reference of Ref.\cite{IMO}.)

In the confinement phase of the present system, 
the only CS fermions $\psi_x$ or the electrons may appear 
as quasiparticles, while in the deconfinement phase, 
the chargeons and fluxons may appear as quasiparticles.
The latter case corresponds to the PFS. Therefore, the PFS is 
characterized  as a deconfinement phase of the gauge dynamics
of the gauge field $(\lambda_x, \;A_{xj})$.\cite{PFS2}
In fact, in the PFS states, 
one may set $V_{xj} = |V_{xj}| U_{xj}$ 
by some constant $V_{xj} \simeq V_0 U_{xj} \simeq V_0$ 
as the first approximation. (We shall calculate $V_0$ later
in Sec.\ref{V0}.)
Then the coupling $J_{xj}V_{xj}$ reduces to the sum of 
chargeon and fluxon hopping terms. The chargeons $\eta_x$ 
just hop in a constant field described by $M_{x+j}M_x^\dagger$,
i.e., in the reduced magnetic field;
\begin{eqnarray}
\Delta B \equiv B^{\rm ex} - B_\phi =
\frac{2\pi \rho}{e}\Big(   \frac{1}{\nu}-2q
\Big) 
=\pm\frac{2 \pi\rho}{ep}.
\label{deltab}
\end{eqnarray}
The fluxons $\phi_x$ hop in a field $W_{x}{W}_{x+j}^\dagger$ 
generated by $\phi_x$ themselves. At low $T$, we expect that
$\phi_x$ form a Bose condensation and one may set 
$\phi_x^\dagger \phi_x = n$ as the first approximation.
Then $W_x \rightarrow 1$ and $\phi_x$ describes just free bosons.
The central problem is to identify the condition when the deconfinement
phase, hence the PFS, is realized in the present system (\ref{J}).

Before studying this problem in detail, let us present some general
considerations. First, we observe that the Lagrangian $L_V$ contains
{\em no} kinetic (i.e., Maxwell) terms of gauge field like
$g^{-2} \sum V^{\dagger}_{xj}V^{\dagger}_{x+j,i}V_{x+i,j}V_{xi}$
(which corresponds to the $(\partial_i A_j - \partial_j A_i)^2$ term 
in the continuum);
that is, the gauge coupling constant $g^2$ is infinite.
Therefore one may suspect that $V_{xj}$ should fluctuate randomly
leading to the confinement phase, so  the PFS does no take place at all.
This argument is valid if the system is a pure gauge theory, 
Lagrangian
of which consists of only a Maxwell term and no couplings to charged 
matter fields.\cite{wilson}
However, our system has couplings to matter fields, chargeons and 
fluxons, $\phi_{x+j}^\dagger V_{xj} \phi_x$ and $\eta_{x+j}^\dagger 
V_{xj} \eta_x$. It is possible that these couplings  may suppress
gauge-field fluctuations at low-energies
as a renormalization effect due to the 
high-momentum and/or high-energy modes of $\phi_x$ and $\eta_x$.
If these factors are efficient enough, the gauge dynamics is
to be realized in the deconfinement phase,
hence the PFS takes place.

Similar question arose for the CSS, and some physicists concluded that
the CSS never takes place in the $U(1)$ gauge theory of
strongly-correlated models like the t-J model.\cite{Nayak}
In our previous paper,\cite{IMO}, we clarified this misunderstanding
by pointing out the renormalization effects by couplings to matter fields.
Also we emphasized there that there exists a counter example 
against the above naive expectation.
For example, the SU(3) gauge theory in $(3+1)$ dimensions 
{\it without} quark fields 
is always in the confinement phase regardless of the magnitude of  
gauge coupling constant $g$.
However, onset of  couplings to quarks, relativistic fermions, 
changes its phase structure drastically.\cite{QCD}
Especially, in quantum chromodynamics (QCD) 
with a sufficiently large number of light quarks 
(the number of quark flavor $N_f \geq 7$),
the deconfinement phase is realized even if the coupling constant
$g_{\rm QCD} \rightarrow \infty$. Furthermore, for $N_f \geq 17$, 
the effective coupling constant is renormalized as
$g_{\rm eff}\rightarrow 0$ at lower energies even for 
$g_{\rm QCD} = \infty$.
One may argue that a similar phenomenon should certainly occur 
in the present gauge system, 
because the chargeons, nonrelativistic fermions, behave
as a set of many species of relativistic fermions. This is because
nonrelativistic fermions have
a Fermi surface (Fermi line in two-dimensions) instead of Fermi points 
of relativistic fermions, and it contains 
a large number of low-energy excitations.
In the following Section, we shall study this possibility
by deriving an effective action of 
the dynamical gauge field $A_{x\mu}$, finding that the PFS is possible 
at sufficiently low temperatures.

\subsection{Repulsive interaction between electrons}
\label{repulsiveint}

To make an explicit and  detailed study of the gauge dynamics, 
one needs to specify the interaction term $H_{\rm int}$ in (\ref{J}).
This interaction between electrons (or CS fermions)
should play a crucial role in the CF picture and the PFS.
Actually, if this term were missing, the system would be
just an ensemble of independent electrons, each electron
occupying degenerate LL's, and no FQHE would be observed.
Generally speaking, appropriate interactions are certainly necessary 
in order that separation phenomena of degrees of freedom take place.

Let us focus on the effects of short-range part
of Coulomb interaction by taking the following 
nearest-neighbor repulsion as $H_{\rm int}$;
\begin{equation}
H_{\rm int}(C^\dagger_xC_x)=H_{\rm int}(\psi^\dagger_x\psi_x)=
g\sum \psi^\dagger_{x+j}\psi_{x+j}\psi^\dagger_x\psi_x,
\label{Hint}
\end{equation}
where $g( > 0)$ is the coupling constant of the Coulomb repulsion.
Since we set $a \simeq \ell$, $g$ is estimated as 
\begin{equation}
g \simeq {e^2 \over \epsilon \ell},
\label{gs}
\end{equation}
where $\epsilon$ is the dielectric constant of materials.
The effects of the long-range part $H^{\rm LR}$ of 
Coulomb repulsion  shall be discussed in Sec.\ref{groundstate}.
By using the relations $\psi^\dagger_{x+j}\psi_{x+j}
\psi^\dagger_x\psi_x
=\eta^\dagger_{x+j}\eta_{x+j}\eta^\dagger_x\eta_x
=\phi^\dagger_{x+j}\phi_{x+j}\phi^\dagger_x\phi_x$ by (\ref{state}),
$H_{\rm int}$ above is rewritten as
\begin{eqnarray}
&&H_{\rm int}
= \sum_{x,j} \Big( g_1 \eta^\dagger_{x+j}\eta_{x+j}\eta^\dagger_x\eta_x
+ g_2 \phi^\dagger_{x+j}\phi_{x+j}\phi^\dagger_x\phi_x \Big),\nonumber\\
&&g_1 + g_2  = g.
\label{g1g2}
\end{eqnarray}
The parameters $g_1$ and $g_2$ shall be determined shortly.
Then we note that this $H_{\rm int}$ has the same form as 
$H_4$ in $L_V$  of (\ref{J}).
Since  we expect that chargeons and fluxons become quasiexcitations
at low energies, we should adjust
their environments so that they behave as freely as possible. 
This self-consistency condition leads us to require
that $H_{\rm int}$ and $H_4$ in $L_V$  of (\ref{J}) 
should cancel out each other,
\begin{eqnarray}
&&H_4 + H_{\rm int} = 0.
\label{selfconsistent}
\end{eqnarray}
Then $g_1$ and $g_2$ are related with $\gamma$ as
\begin{eqnarray}
&&g_1=\frac{1}{2ma^2\gamma^2},\ g_2 =-\frac{\gamma^2}{2ma^2}.
\label{h4hint}
\end{eqnarray}
From $g_1 + g_2 = g$, $\gamma$ satisfies
\begin{equation}
\frac{1}{2ma^2\gamma^2} -\frac{\gamma^2}{2ma^2} 
\simeq {e^2 \over \epsilon \ell}.
\label{gamma}
\end{equation}
The relation (\ref{gamma}) is suggestive, expressing the 
relation between two energy scales, one is the inter-LL gap
for electrons (the cyclotron frequency), 
\begin{eqnarray}
\omega_B = \frac{e B^{\rm ex}}{m},
\end{eqnarray}
and another
is the short-range Coulomb energy $e^2/(\epsilon \ell) 
(\propto \sqrt{B^{\rm ex}})$ 
for chargeons and fluxons. In fact, 
(\ref{gamma}) is rewritten with $a \simeq \ell$ as
\begin{equation}
\frac{1}{2ma^2}(\frac{1}{\gamma^2} -\gamma^2)
\simeq \frac{\omega_B}{2}(\frac{1}{\gamma^2} -\gamma^2) 
\simeq {e^2 \over \epsilon \ell}.
\label{gamma2}
\end{equation}
Thus the dimensionless constant  $\gamma^{-2} -\gamma^2$
is a reduction factor to reduce  $\omega_B$ down to 
$e^2 /( \epsilon \ell)$.  
In the experiment, the effective electron mass $m$, 
the dielectric constant $\epsilon$, and the
magnetic length $\ell$ are all given, so 
 $\gamma$ is determined from (\ref{gamma}) 
 by these parameters.\cite{HCboson,Definition}

\section{Effective Gauge Theory and PFS}
\setcounter{equation}{0}

In this Section, we derive the effective lattice gauge theory
of the gauge field $A_{x\mu}$ by the hopping expansion
over  $\phi_x$ and $\eta_x$, and study its phase structure.
In the effective gauge theory, we shall ignore  fluctuations
of the absolute value $|V_{xj}|$ since they are massive,
and focus on its U(1) phase part $U_{xj}$. We shall also see
the fluctuations of
$\lambda_x = A_{x0}$ are massive and can be ignored.
Thus the effective action is a function of $U_{xj}$'s. 
As the lattice regularization is used in the present study, 
we can study the problem nonperturbatively. 
This is in sharp contrast to the most of other studies of the CS gauge
theory.

\subsection{Estimation of $V_0  \equiv \langle |V_{xj}| \rangle$}
\label{V0}

We start with writing the spatial component of the  
gauge field, $V_{xj}$, in the polar coodinate as
\begin{eqnarray} 
V_{xj} &=& |V_{xj}|U_{xj},\ U_{xj} = \exp(iA_{xj}) \in {\rm U(1)}.
\end{eqnarray}
Let us first estimate the expectaiton value of the amplitude,
 $V_0 \equiv \langle |V_{xj}| \rangle$
by  MFT, which is obtained by setting $V_{xj}=V_0$ and 
$\lambda_x=0$ in the Lagrangian $L_V$ of (\ref{J}).
This MFT is jusitified for studying the low-energy physics a 
posteriori, 
because fluctuations of both  $|V_{xj}|$ and $\lambda_x$ 
are shown to be massive and irrelevant to the low-energy physics.
On the other hand, the phase part, $A_{xj}$, may be massless and should be 
treated carefully.
Estimation of $V_0$ is further simplified by putting the fluxon variables
as $\phi_x=\sqrt{n}$, since one expects
Bose condensation of fluxons at low $T$. 
In the bosonic CS gauge theory of FQHE, Bose condensation actually
takes place and the Bose-condensed ground state describes the Laughlin state
as is shown in Ref.\cite{CB}. The Lagrangian $L_{V_0}$ of the MFT 
then takes the following form;
\begin{eqnarray}
L_{V_0}&=&-\sum_x \eta_x^\dagger(\partial_\tau-\mu_\eta)\eta_x
\nonumber\\
&+&\frac{1}{2ma^2} \sum_{x,j}\Big[-V_0^2+V_0
\Big(\gamma n  +\frac{1}{\gamma}
\eta^\dagger_{x+j}\eta_x+ {\rm H.c.}\Big)\Big].\nonumber\\
\label{LMFT}
\end{eqnarray} 
A typical behavior of $V_0$ was plotted in Fig.1
of Ref.\cite{PFS1}. It decreases as $T$ increases.
We expect $V_0$ vanishes at certain temperature  $T_{V_0}$. 
Near $T_{V_0}$, the MFT (\ref{LMFT})
is not reliable since Bose condensation would disappear.
Instead, an expansion in term of small $V_0$ is possible.\cite{PFS1}
We shall calculate $T_{V_0}$ in (\ref{TV0value}) later.

The small fluctuations of $|V_{xj}|$ around $V_0$ are described
by inserting $V_{xj} = V_0 + v_{xj}, v_{xj} \in R$ to $V_0$ in $L_{V_0}$
and expand $L_{V_0}$ up to $O(v_{xj}^2)$.  The squared mass
of $v_{xj}$ is positive, hence these fluctuations are irrelevant
at low energies. 
 
\subsection{Hopping expansion and effective gauge theory} 
  
In the gauge theory of nonrelativistic fermions, the time component of
gauge field is often screened (getting massive) and becomes irrelevant 
to the low-energy physics. This is exploited, for example, in the study 
of CS gauge theory of CF at $\nu = 1/2$
by Halperin, Lee and Read.\cite{HLR}
Let us examine this problem.

From Sec.\ref{V0}, we are allowed to set
\begin{eqnarray}
V_{xj}&=&V_0 U_{xj},\ U_{xj} = \exp(iA_{xj}) \in {\rm U(1)}
\end{eqnarray} 
below $T_{V_0}$. 
Thus we are to study the dynamics of the U(1) gauge 
field $U_{xj}$ and $\lambda_x$.
To this end, we obtain an effective action of $U_{xj}$ 
and $\lambda_x$ by
using the hopping expansion over $\phi_x$ and $\eta_x$,
which is an expansion in powers of $V_0$ or equivalently of 
$U_{xj}$.
This expansion is a nonperturbative expansion w.r.t. $A_{xj}$
that is legitimate for small expectation values of $U_{xj}$.
It is especially useful to study the phase transition
for which $\langle U_{xj} \rangle$ may be regarded as an 
order parameter with a continuous change.

To be explicit, we use the so-called temporal gauge.
At $T=0$, one can set $\lambda_x(\tau)=0$.
However, at finite $T$, one degree of freedom of $\lambda_x(\tau)$
should survive as an independent variable.\cite{wilson}
We make the eigen-frequency expansion,
\begin{eqnarray}
\lambda_x(\tau)= \sum_{n \in Z} \lambda_{x,n} \exp(i\omega_n \tau),\
\omega_n = \frac{2\pi n}{\beta},
\end{eqnarray}
and let the zero mode $\theta_x$,
\begin{eqnarray}
\theta_x &\equiv& \lambda_{x,0} = \int_0^\beta d\tau \lambda_x(\tau)
\end{eqnarray} 
as the remaining integration variable. (This $\theta_x$ should not
be confused with the azimuthal angle of $x$.)
All the other modes are set zero; $\lambda_{x,n}\ (n \neq 0) =0$. 

In the hopping expansion, we need the on-site  
propagators of $\phi_x$ and $\eta_x$,
$\langle \phi_x(\tau_1)\phi^\dagger_y(\tau_2)\rangle_0,\
\langle \eta_x(\tau_1)\eta^\dagger_y(\tau_2)\rangle_0$ 
determined from the following 
Lagrangian $L_0$ obtained from $L_V$ by setting $V_{xj} = 0$;
\begin{eqnarray}
L_0&=&-\sum_x 
\eta_x^\dagger(\partial_\tau+i\beta^{-1}\theta_x-\mu_\eta)\eta_x
\nonumber\\
&&- \sum_x \phi_x^\dagger(\partial_\tau-i\beta^{-1}\theta_x-\mu_\phi)\phi_x. 
\end{eqnarray}
Calculations are straightforward, and we obtain
\begin{eqnarray}
&&\langle \eta_x(\tau_1)\eta^\dagger_y(\tau_2)\rangle_0 =\delta_{xy}
{e^{\mu_\eta(\tau_1-\tau_2)} \over 1+e^{\beta\mu_\eta-i\theta_x}} 
\nonumber \\
&& \times \Big[e^{-i\theta_x(\tau_1-\tau_2)/\beta}\theta(\tau_1-\tau_2)
\nonumber\\
&&- e^{\beta\mu_\eta-i\theta_x-i\theta_x (\tau_1-\tau_2)/\beta}
\theta(\tau_2-\tau_1) \Big],
\label{propa1}
\end{eqnarray}
\begin{eqnarray}
&&\langle \phi_x(\tau_1)\phi^\dagger_y(\tau_2)\rangle_0 =\delta_{xy}
{e^{\mu_\phi(\tau_1-\tau_2)} \over 1-e^{\beta\mu_\phi+i\theta_x}} 
\nonumber \\
&& \times \Big[e^{i\theta_x(\tau_1-\tau_2)/\beta}\theta(\tau_1-\tau_2)
\nonumber\\
&&+ e^{\beta\mu_\phi+i\theta_x+i\theta_x (\tau_1-\tau_2)/\beta}
\theta(\tau_2-\tau_1) \Big],
\label{propa2}
\end{eqnarray}
where $\theta(\tau)$ is the step function. 
The chemical potentials are determined by
\begin{eqnarray}
\langle \eta^\dagger_x \eta_x \rangle &=& \lim_{\epsilon \rightarrow +0}
\langle \eta^\dagger_x(\tau+\epsilon)\eta_x(\tau)\rangle_0  = 
{e^{\beta\mu_\eta} \over 1+e^{\beta\mu_\eta}} =n, \nonumber  \\
\langle \phi^\dagger_x \phi_x \rangle &=& \lim_{\epsilon \rightarrow +0}
\langle \phi^\dagger_x(\tau+\epsilon)\phi_x(\tau)\rangle_0=
{e^{\beta\mu_\phi} \over 1-e^{\beta\mu_\phi}} = n.
\label{rho}
\end{eqnarray}

By expanding the integrand of $Z$ of (\ref{J}) in powers of
$V_0$, and by using the propagators, (\ref{propa1}) and (\ref{propa2}), 
the effective action of the gauge field  
$A_{\rm eff}(\theta,U)$ is obtained, which is a kind of 
GL theory for the gauge field $(\theta,U)$;\cite{hopping} 
\begin{eqnarray}
 Z&=&\int [dU][d\theta]
\exp \Big(A_{\rm eff}(\theta,U) \Big), \nonumber   \\
 A_{\rm eff}(\theta,U)&=&A_0(\theta)+A_2(\theta,U)
+{\rm O}(V_0^4).
\label{Z2}
\end{eqnarray}
In the leading order  O$(V_0^0)$ of the expansion,
the effective action, $A_0(\theta)$ of $\theta_x$, is obtained as
\begin{eqnarray}
A_0(\theta)&=&
\sum_x\ln F(\theta_x),\nonumber\\
F(\theta_x) &\equiv& {1+e^{\beta\mu_\eta-i\theta_x} 
\over 1-e^{\beta\mu_\phi+i\theta_x}} 
=  {1+\displaystyle{\frac{n}{1-n}}e^{-i\theta_x} \over 
1-\displaystyle{\frac{n}{1 + n}} e^{i\theta_x}}.
\label{L0}
\end{eqnarray}
For the decoupled partition function at each $x$ for $A_0$, we have
\begin{eqnarray}
\int_0^{2\pi} \frac{d \theta_x}{2\pi} F (\theta_x) 
&=& 1 + e^{\beta(\mu_\phi + \mu_\eta)}
= \sum_{n_x=0}^1 e^{\beta(\mu_\phi + \mu_\eta) n_x},\nonumber\\
\label{L02}
\end{eqnarray}
as expected from the constraint. (Here $n_x = 
\eta^\dagger_x \eta_x = \phi^\dagger_x \phi_x$ are the particle
numbers.)
We note that Re $F(\theta_x)$ is an even funciton 
of $\theta_x$ and has the maximum at $\theta_x = 0$ (mod $2\pi$), 
while Im $F(\theta_x)$ is an odd funciton of $\theta_x$.
 
Similarly, the second-order calculation of the hopping 
expansion gives rise to
\begin{eqnarray}
&A_2(\theta,& U) =-\beta\sum_{x,j}\frac{V_0^2}{2ma^2}
+\sum_{x,j} \int^\beta_0 d\tau_1 \int^\beta_0 d\tau_2  \nonumber\\
&& \ \ \ \ \ \ \ \times C(\theta_x, \theta_{x+j; }\tau_1,\tau_2)
U^\dagger_{xj}(\tau_1)U_{xj}(\tau_2), \nonumber  \\
&C(\theta_x,& \theta_{x+j; }\tau_1,\tau_2) =
V_0^2\Big[ \Big({\gamma \over 2ma^2}\Big)^2 \nonumber  \\
&&\times
\langle \phi_{x+j}W_xW^\dagger_{x+j}\phi^\dagger_x(\tau_1)\cdot
\phi_{x}W_{x+j}W^\dagger_{x}\phi^\dagger_{x+j}(\tau_2)\rangle_0 
\nonumber \\
&&+\Big({1\over 2\gamma ma^2}\Big)^2 \langle
\eta^\dagger_{x+j}\eta_x(\tau_1)
\eta^\dagger_x\eta_{x+j}(\tau_2)\rangle_0\Big]\nonumber\\
&&\equiv {V^2_0 \over 4m^2a^4} 
f(\theta_x,\theta_{x+j};\tau_1-\tau_2),
\label{S2f}
\end{eqnarray}
where $f(\theta_x,\theta_{x+j};\tau_1-\tau_2)$ is a certain complicated 
function of $\theta_x$ and $\theta_{x+j}$.
In particular, for $\tau_1-\tau_2 \sim 0$, we have
\begin{eqnarray}
&&f(\theta_x,\theta_{x+j};\tau_1-\tau_2)\nonumber\\
&\simeq& 
\Big[
\theta(\tau_2 - \tau_1)e^{-i\theta_x} 
+\theta(\tau_1 - \tau_2)e^{-i\theta_{x+j}} 
\Big]\nonumber\\
&\times&
\Big[
\frac{\gamma^{-2}
e^{\beta\mu_\eta}
}
{
(1+e^{\beta\mu_\eta-i\theta_x})
(1+e^{\beta\mu_\eta-i\theta_{x+j}})
}\nonumber\\
&+&
\frac{\gamma^2
e^{\beta\mu_\phi}}
{(1-e^{\beta\mu_\phi+i\theta_x})
(1-e^{\beta\mu_\phi+i\theta_{x+j}})}
\Big].
\label{f}
\end{eqnarray}

\subsection{Integration over zero modes $\theta_x$} 

Let us make a rough estimation of $\theta_x$-integration in (\ref{Z2})
by using   (\ref{L0}), (\ref{S2f}), and (\ref{f}).
We shall obtain a simple result that $\theta_x$  can be treated
as small fluctuations around the minimum $\theta_x = 0$.
To see this, we start  by replacing 
$U_{xj}$-dependent part in $A_2(\theta,U)$
by its average. Explicitly, we start with the following 
general expression of $Z$  
up to O$(U^2)$ after the $\theta$-integration;
\begin{eqnarray}
Z &=& \int [dU] \exp(A_2(U)),\nonumber\\ 
A_2(U) &=&  \int d\tau_1 d\tau_2 C_U(\tau_1-\tau_2)\; \sum_{x,j} 
U^\dagger_{xj}(\tau_1) U_{xj}(\tau_2),
\label{SU}
\end{eqnarray}
where $C_U$ depends on $T$ and $n$, and it is shown to be positive 
for $ \tau_1-\tau_2 \sim 0$ by (\ref{f}).   
From this result, configurations like 
$U^\dagger_{xj}(\tau_1)U_{xj}(\tau_2)\sim 1$ dominate the functional 
integral at least for $(\tau_1-\tau_2) \sim 0$.
Then one can assume 
\begin{equation}
\langle U^\dagger_{xj}(\tau_1)U_{xj}(\tau_2) \rangle \propto
e^{-m_0 |\tau_1-\tau_2|}\ {\rm or}\  \sim |\tau_1-\tau_2|^{-r},\ r > 0. 
\label{corrU}
\end{equation}
By substituting the correlator (\ref{corrU}) into (\ref{S2f}) and by the 
decoupling, the $\tau$-integrals in $A_2(U)$ takes the form,
\begin{eqnarray}
&&\int d\tau_1d\tau_2 f(\theta_x,\theta_{x+j};\tau_1-\tau_2)
\langle U^\dagger_{xj}(\tau_1)U_{xj}(\tau_2) \rangle \nonumber   \\
&=& -b (\nabla_j \theta_x)^2,
\label{kinetic}
\end{eqnarray}
with a positive constant $b$.
From (\ref{L0}) and (\ref{kinetic}), the general form  of 
the effective action for $\theta_x$ is fixed as
\begin{eqnarray}
A_2(\theta) &=& -\sum_{x}\Big[b\sum_j(\nabla_j \theta_x)^2
+\ln F(\theta_x)\Big].
\end{eqnarray}
The first kinetic term generates correlations among $\theta_x$ and
disfavors configurations of independent $\theta_x$'s.
In such a field theory (having infinite degrees of freedom) the limit
$b \rightarrow 0$ is singular, and one obtains
qualitatively correct results for $b > 0$ by expanding $\theta_x$
around $\theta_x = 0$, the minimum of $\ln F(\theta_x)$, 
up to $O(\theta_x^2)$. In other words,  
$\lambda_x=\theta_x/\beta=0$ in the 
ground state and their excitations are massive.
This implies that the effects of $\lambda_x$ are screened by 
fluctuations (radiative corrections)  of chargeons and fluxons, 
and the constraint (\ref{const}) becomes irrelevant at low energies.
This nonperturbative observation supports the ``weak-coupling" 
perturbative calculation in gauge theories of nonrelativistic fermions
which is often used in random-phase 
approximations, etc.
An intuitive picture of the above screening phenomenon at long wavelength
is supplied by block-spin-type renormalization-group (RG) 
transformations.
At first, numbers of  chargeons and fluxons at each lattice site 
are well-defined quantities.
However, through block-spin RG transformations, the number of particles 
at each site of the transformed larger lattice becomes ambiguous.
As a result, the local constraint (\ref{const}) tends to irrelevant
at long wavelengthes.

\subsection{Confinement-deconfinement phase transition}

The above result
that the ground state of $\theta_x$ is  $\theta_x=0$ 
and its excitations are massive
leads that the question whether the PFS takes place or not
is determined just by the dynamics of the transverse gauge field $U_{xj}$.
The PFS corresponds to the
deconfinement phase of $U_{xj}$.

Thus, to evaluate $A_2$ of (\ref{S2f}), 
we simply put $\theta_x=0$ in $f(\theta_x,\theta_{x+j};\tau_1
-\tau_2)$. Then by using (\ref{rho}), we have  
$f(0,0;\tau_1-\tau_2)=\gamma^{-2}n(1-n)+\gamma^2n(1+n)$. So
$A_2$  reduces to\cite{correction} 
\begin{eqnarray}
&&A_2(\theta_x=0,V_0U_{xj})=-\beta\sum_{x,j}\frac{V_0^2}{2ma^2}
  \nonumber   \\
&&\hspace{0.5cm}   +D_2 V_0^2 \sum_{xj}
\int^\beta_0d\tau_1\int^\beta_0d\tau_2
U^\dagger_{xj}(\tau_1)U_{xj}(\tau_2) \nonumber  \\
&&\hspace{0.5cm} = V^2_0\sum_{x,j}\Big[-\frac{\beta}{2ma^2}+D_2
 \beta^2 U^\dagger_{xj,0}U_{xj,0}
\Big],\nonumber\\
&&D_2 \equiv \frac{1}{4m^2a^4}\Big(\frac{n(1-n)}{\gamma^{2}}
+\gamma^{2}n(1+n)\Big),
\label{S2}
\end{eqnarray}
where we have introduced Fourier decomposition of $U_{xj}(\tau)$,
\begin{eqnarray}
&& U_{xj}(\tau)=\sum_{n \in Z} U_{xj,n}\ e^{i\omega_n\tau},\
\omega_n \equiv \frac{2\pi n}{ \beta}, \nonumber  \\
&& \sum_n U^\dagger_{xj,n}U_{xj,n+m}=\delta_{m0},
\label{Fourier}
\end{eqnarray}
where the second line represents 
$U^\dagger_{xj}(\tau) U_{xj}(\tau) =1$.
The  higher-order terms of $A_{\rm eff}$ in 
the hopping expansion can be calculated in a 
straightforward manner.
There appear plaquette terms (i.e., magnetic terms) like
\begin{eqnarray}
A_{\rm plaq}&=& C_{\rm plaq} 
\sum_{x}U_{x2,0}U_{x+2,1,0}U^\dagger_{x+1,2,0}U^\dagger_{x1,0}
+ {\rm H.c.}
\label{plt}
\end{eqnarray}
Their coefficient $C_{\rm plaq} $ 
is positive and also getting large at low $T$.

From (\ref{S2}) and (\ref{plt}), the static-modes of the transverse 
gauge field $U_{xj,0}$ are enhanced at large $\beta$ and 
nonvanishing $V_0$, i.e., $\langle U_{xj,0}\rangle \equiv U_0 \sim 1$ 
(up to irrelevant pure-gauge freedoms).
Explicitly, we estimate that $U_{xj,0}$ has nonvanishing 
expectation value $U_0$ 
when the coefficient of $|U_{xj,0}|^2$ in $A_2$ is larger than
unity,
\begin{eqnarray}
D_2 V_0^2 \beta^2 > 1.
\label{PFScondition}
\end{eqnarray}
Since $V_0$ is a decreasing function of $T$, 
this happens at lower $T$'s.
On the other hand, the other oscillating modes $U_{xj,n\neq 0}$
are strongly suppressed by (\ref{Fourier}),
i.e., $\sum_{n \neq 0}|U_{xj,n}|^2 \ll 1$.
This implies that, at low $T$ where the condition (\ref{PFScondition})
holds, the gauge dynamics is in the deconfinement phase, where
$\langle U_{xj}(\tau) \rangle \simeq U_0$ (up to gauge freedom) and 
the fluctuations $\Delta A_{xj}$ are small.
Therefore, PFS takes place at low $T$ such that (\ref{PFScondition})
holds\cite{CD}.
The quasiexcitations there are the chargeons $\eta_x$, fluxons
$\phi_x$, and the gauge bosons $A_{xj}$.
The transition temperature $T_{\rm PFS}$ can be estimated 
from the ``GL theory" (\ref{S2}) by setting the coefficient
of $|U_{xj,0}|^2$  unity; 
\begin{eqnarray}
&&D_2 V_0^2 \beta^2 |_{T = T_{\rm PFS}}  \nonumber\\
&=&  \frac{V_0^2(T_{\rm PFS})}{4m^2a^4 
k_{\rm B}^2 T^{2}_{\rm PFS}} \Big(\frac{n(1-n)}{\gamma^{2}}
+\gamma^{2}n(1+n)\Big)
\simeq 1.
\label{TPFS}
\end{eqnarray}

The analysis using the method by Polyakov and Susskind\cite{CD} 
predicts that the phase transition at $T_{\rm PFS}$ is smooth as in 
CSS,\cite{CSS,IMO} so our hopping expansion of $A_{\rm eff}$ 
in powers of $V_0 U_{xj}$ is justified a posteriori.\cite{hopping} 
Concerning to the order of  
the CDPT of the present system,  we remind that
the usual lattce gauge theory on a 3D spatial lattice
exhibits a second-order CDPT, while the gauge theory
on a 2D spatial lattice exhibit a CDPT of the 
Kosterlitz-Thouless (KT) type.\cite{CD} 
However, the present effective lattice
gauge theory (\ref{S2}) on a 2D lattice 
has a stronger correlations along $\tau$-direction than 
the usual coupling $|\partial_\tau U_{xj}|^2$, so the 
CDPT may be of second order instead of the KT type.
Further study is required to clarify this point.

For comparison, let us list up
the temperature regions in which the deconfinement phases take place
for various models;
\begin{eqnarray}
\begin{tabular}
{|c|c|}  \hline
      Model             &  {Region of deconfinement}
      \\ \hline
QED,\ QCD($7 \leq N_f$)  & $ 0 \leq T$      
      \\ \hline
QCD($0 \leq N_f \leq 7$)  & $ 0 < T_{\rm CD} < T$      
      \\ \hline
t-J Model & $ 0 \leq T < T_{\rm CSS}$      
      \\ \hline
FQH System  & $ 0 \leq T < T_{\rm PFS}$      
\\ \hline
\end{tabular}
\label{RegionOfT}
\end{eqnarray} 
One may feel it strange that the deconfinement region is at high $T$
for QED and QCD, while it is low $T$ for the t-J model and the present
FQH system. From Ref.\cite{CD}, the condition of deconfinement is
estimated  as $\beta g^2_{\rm G} < 1$ for a gauge theory with
a gauge coupling constant $g_{\rm G}$. When $g_{\rm G}$ is a $T$-independent
constant as in QED and QCD, the deconfinement occurs at high $T$.
However, in the strongly-correlated electron systems, the gauge theory
in question is an effective theory obtained by integrating out the 
``electron" degrees of freedom, and the resulting gauge coupling may be
$T$-dependent. In fact, we estimated $g_{\rm G}^2 \propto 
T^{3}$ for the t-J model,\cite{CSS} which implies the result 
(\ref{RegionOfT}). By repeating the same argument as Ref.\cite{CSS},
we obtain the same result $g_{\rm G}^2 \propto 
T^{3}$ for the present gauge model, which is consistent with
the condition (\ref{PFScondition}).

\subsection{Numerical results}
\label{numericalresults}

Let us estimate $T_{\rm PFS}(\nu)$ 
numerically  by using (\ref{TPFS}) 
with $V_0(T)$ determined by (\ref{LMFT}). 
We consider the following choice of the Coulomb coupling;
\begin{eqnarray}
g = (0.1 \sim 1) \times \frac{e^2}{\epsilon \ell},
\label{g}
\end{eqnarray}
and the parameters,
\begin{eqnarray}
&& a = \ell,\ \ B^{\rm ex}=10[{\rm T}], \nonumber\\
&& m=0.067 \times m_{\rm electron},\ \  \epsilon=13.
\label{parameters}
\end{eqnarray} 
The parameter $\gamma$ is $\nu$-independent and 
calculated by (\ref{gamma}) as 
\begin{eqnarray}
\gamma=0.96 \sim 0.69,
\label{parameters2}
\end{eqnarray}
for (\ref{g}).

\begin{figure} 
  \begin{picture}(200,140) 
    \put(0,0){\epsfxsize 200pt \epsfbox{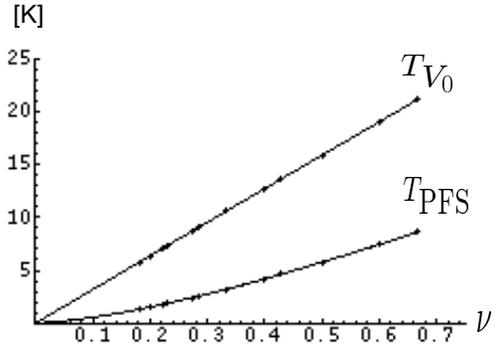}} 
  \end{picture} 
  \caption{$T_{\rm PFS}$ and $T_{V_0}$ versus $\nu$ 
  for  $g = 0.1 e^2/(\epsilon \ell)$ calculated by using
  (\ref{TPFS}) and (\ref{TV0value}).
  The dots represent the cases of $\nu =$ 2/3, 3/5, 1/2, 3/7, 2/5, 
  1/3, 2/7, 3/11, 3/13, 2/9, 1/5, 2/11  (Same in Fig.\ref{T10},
  \ref{METAPHI01}, \ref{METAPHI10}). 
  }
\label{T01}
\end{figure}

\begin{figure} 
  \begin{picture}(200,140) 
    \put(0,0){\epsfxsize 200pt \epsfbox{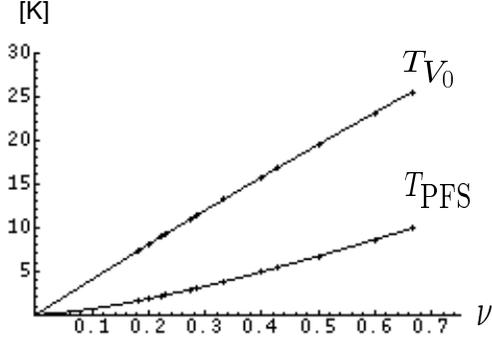}} 
  \end{picture} 
  \caption{$T_{\rm PFS}$ and $T_{V_0}$ versus $\nu$ 
  for  $g =  e^2/(\epsilon \ell).$
  }
\label{T10}
\end{figure}

We plot $T_{\rm PFS}(\nu)$  at  various fillings 
$\nu = p/(2pq \pm 1)$ for $g = 0.1 e^2/(\epsilon \ell)$
in Fig.\ref{T01} and for $g = e^2/(\epsilon \ell)$ in Fig.\ref{T10}.
At $\nu = 1/2$,  
\begin{eqnarray}
T_{\rm PFS}(1/2) &=& 5.7 \sim 6.7 {\rm K}.
\label{TPFS2}
\end{eqnarray} 
$T_{\rm PFS}(\nu)$ seems consistent with the 
experiments.\cite{PFSexp} We note that the highest temperature  at which 
FQHE  is observed is $T_{\rm BC} ( < T_{\rm PFS})$  
since Bose condensation of fluxons is necessary for FQHE
(See Sec.\ref{transport} for details).

Let us estimate the effect of phase fluctuations of $U_{xj}$
upon PFS. For this purpose, we compare $T_{\rm PFS}$ with 
its ``mean-field" value $T_{V_0}$.
$T_{V_0}$ is defined by the temperature at which $V_0$ starts to 
develop and is calculated by setting 
the coefficient of the $V_0^2$ term in
 $A_{\rm eff}$ with $|U_{xj,0}| =1$ to vanish, i.e.,
  $-\beta/(2ma^2) + \beta^2 D_2 = 0$ as  
\begin{eqnarray}
T_{V_0} &=&    \frac{1}{2ma^2 k_{\rm B}}
\Big(\frac{n(1-n)}{\gamma^{2}} + \gamma^2 n(1+n) \Big). 
\label{TV0value}
\end{eqnarray} 
This $T_{V_0}(\nu)$ is also plotted in Fig.\ref{T01},\ref{T10}.
At $\nu = 1/2$,
\begin{eqnarray}
 T_{V_0}(1/2) = 16 \sim 19 {\rm K},
\label{TV01/2}
\end{eqnarray}
which is about 3 times larger than $T_{\rm PFS}(1/2)$ of (\ref{TPFS2}).
This demonstrates the importance of fluctuations of
the gauge-field $A_{xj}$ in PFS, which
reduce the critical temperature of PFS significantly
from $T_{V_0}$ down to $T_{\rm PFS}$. The similar large
effect of gauge field was found also in CSS.\cite{IMS}

\begin{figure} 
  \begin{picture}(200,140) 
    \put(0,0){\epsfxsize 200pt \epsfbox{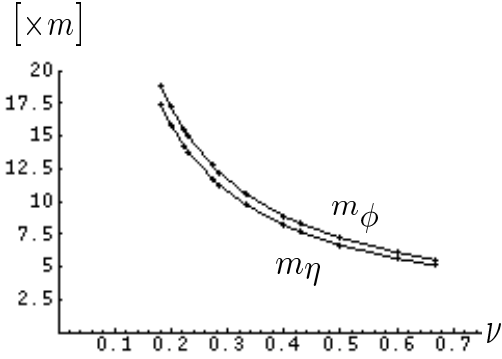}} 
  \end{picture} 
  \caption{$m_\eta, m_\phi$ for $g = 0.1 e^2/(\epsilon \ell)$ 
  calculated  by using (\ref{effectivemass}).
  The curves are roughly fitted by
  $m_\eta \sim 3.2 \nu^{-1},$\ $m_\phi \sim 3.5\nu^{-1}$. }
\label{METAPHI01}
\end{figure}

\begin{figure} 
  \begin{picture}(200,140) 
    \put(0,0){\epsfxsize 200pt \epsfbox{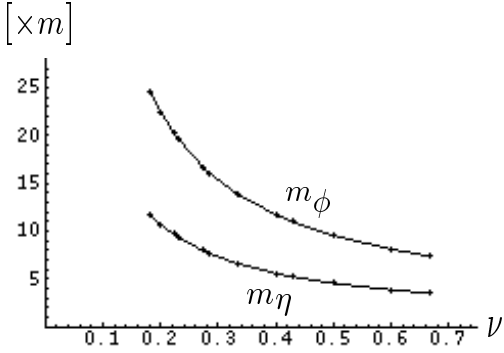}} 
  \end{picture} 
  \caption{$m_\eta,m_\phi$ for $g =  e^2/(\epsilon \ell)$. 
  }
\label{METAPHI10}
\end{figure}

Let us next consider the masses of chargeon and fluxon, 
$m_{\eta}(\nu)$ and $m_{\phi}(\nu)$. 
To calculate $m_{\eta}$ and $m_{\phi}$,
one may set $V_{xj} \rightarrow V_0$ in the $J_{xj}^\dagger V_{xj}$ term 
in Lagrangian (\ref{J}) at low $T$. Then this term gives rise to the 
hopping terms of $\eta_x$ and $\phi_x$, from which one obtains 
\begin{eqnarray}
m_{\eta} &=& \frac{\gamma}{V_0} m,\ m_{\phi} = \frac{1}{\gamma V_0} m. 
\label{effectivemass}
\end{eqnarray}
Their values for (\ref{g}) and (\ref{parameters}) are plotted in 
Fig.\ref{METAPHI01},\ref{METAPHI10}, 
which show  $m_\eta, m_\phi > m$.
At $\nu = 1/2$,
\begin{eqnarray}
m_{\eta}(1/2) &=&   (6.6 \sim 4.5 ) m,\ \
m_{\phi}(1/2) = (7.2 \sim 9.5 ) m. \nonumber\\
\label{CFmass}
\end{eqnarray}
Experimentally, the mass of CF, $m_{\rm CF}$, is determined by 
equating the observed activation energy $E_{\rm ac}$ 
in the resistivity $\rho_{xx} \propto \exp(-E_{\rm ac}/2k_{\rm B}T)$ 
to be an energy
gap $\omega_{\Delta B} \equiv e \Delta B/m_{\rm CF}$ 
between the lowest and the next LL formed by
the residual magnetic field $\Delta B$
felt by CFs, i.e., $m_{\rm CF} =   e \Delta B/E_{\rm ac}$. 
In our theory, chargeons just describe Jain's CFs, so one has
$m_{\eta} = m_{\rm CF}$. The estimate  (\ref{CFmass}) of $m_\eta$
in Fig.\ref{METAPHI01},\ref{METAPHI10}
seems consistent with the experiments \cite{PFSexp}
apart from the region near $\nu = 1/2$ where the effect of coupling 
of gauge field to (almost) massless fermions may be large.\cite{HLR,IMO1}
  
One may argue that the scale of the CF mass, $m_{\eta}$,
should be set by the Coulomb energy, and not by 
the electron band mass $m$. For example, even for
the limit $m \rightarrow 0$ at which the energy gap between neighboring
LL's of electrons diverges, the CF should have a finite mass.
The expressions of $m_{\eta, \phi}$ in  (\ref{CFmass}) are
free  from this criticism. 
They hold just for the physical value of 
the electron band mass $m$, and
do not imply $m_{\eta, \phi}$ are proportional to $m$.
(The ratios $m_{\eta, \phi}/m$ vary as $m$ varies.)
In our approach, we argued  in Sec.\ref{repulsiveint} that 
  the residual interactions $H_{4}$ of 
$\eta_x$ and $\phi_x$, which involves $m$, and the short-range 
Coulomb interaction $H_{\rm int}$ should relate as (\ref{h4hint}).
Thus the formula above may be interpreted as 
$m_{\eta, \phi} \sim e^2/(\epsilon \ell)$.

In this section, we studied the phase structure of the effective
gauge theory, which is defined on a lattice.
One may wonder whether the PFS and the CDPT studied on the lattice model
 survive in the ``continuum limit".
The CDPT at finite $T$  was first discovered
by Polyakov and Susskind\cite{CD} in lattice gauge theory.
After that, more detailed investigations, including numerical studies
and renormalization-group (RG) analyses, 
confirm the existence of this  CDPT in the continuum. 
The lattice models are regarded there as the effective models of 
RG, and the transition tepmerature is a RG-invariant quantity.
Moreover, recent studies on gauge theories in the continuum spacetime,
which is closely related to the
present gauge system, support the existence of 
CDPT at finite $T$. 
For example, in Ref.\cite{kleinert}, the $U(1)$ gauge theory 
in $(2+1)$ dimensions coupled with multi-fermions is studied 
by the RG method.
There a nontrivial infrared fixed point is found.
This result indicates that a deconfinement phase is realized at low $T$.
In Ref.\cite{antonov}, finite$-T$ properties of the compact $U(1)$
gauge theory is studied by using the Georgi-Glashow model in
$(2+1)$ dimensions.
From the instanton calculation, it is shown that a CDPT occurs 
at finite $T$ as predicted by the lattice gauge theory.\cite{CD}
These facts support that our results of
the PFS and CDPT obtained in this section remain valid even in the
``continuum limit".

\section{Ground State, Low-Energy Excitaitons, EM Transport}
\setcounter{equation}{0}
\label{GS}

\subsection{Ground state}
\label{groundstate}

In the previous section, we showed that the PFS takes place at low 
$T < T_{\rm PFS}$.
In the PFS states, the chargeons and fluxons are interacting only
weakly through the gauge field $A_{xj}$. In this perturbative phase of
gauge dynamics, one  can return to the continuum notation.
(Recall that the lattice regularization was essential for the 
strong-coupling phase of gauge dynamics.)
Let us study the ground state by putting $V_{xj}=V_0$.
Due to the separation pheomenon PFS, the Hamiltonian may be
separated effectively into the chargeon part $H_\eta$ 
and the fluxon part $H_\phi$,
\begin{eqnarray}
H_{\rm eff} = H_\eta + H_\phi,
\end{eqnarray}
where we have safely neglected 
perturbative mixing effects mediated by $A_{xj}$.
Then the ground state of electrons $|G\rangle_C$
 is given by a direct product of the ground state of $H_\eta$,
 $|G\rangle_\eta$, and the ground state of $H_\phi$,
 $|G\rangle_\phi$, i.e., 
\begin{eqnarray}
|G\rangle_C=|G\rangle_\eta |G\rangle_\phi.
\end{eqnarray}
Thus the electron wave function is expressed in the continuum 
notation as
\begin{eqnarray}
&&\Psi_e(x_1,\cdots,x_N) \equiv
{}_C\langle 0 |C(x_1)\cdots C(x_N)|G\rangle_C  \nonumber  \\
&&\ \ \ \ \ \ =\Psi_\eta(x_1,\cdots,x_N)\cdot 
\tilde{\Psi}_\phi(x_1,\cdots,x_N),\nonumber\\
&&\Psi_\eta(x_1,\cdots,x_N) \equiv {}_{\eta}\langle 0 |\eta(x_1)
\cdots \eta(x_N)|G\rangle_\eta, \nonumber\\
&&\tilde{\Psi}_\phi(x_1,\cdots,x_N) \equiv 
{}_{\phi}\langle 0 |\phi(x_1)\cdots 
\phi(x_N) \times \nonumber\\
&&\ \ \ \ \ \ \ 
 \prod_k \exp [2iq \int d^2y \theta(x_k-y)\phi^\dagger(y)\phi(y)]
|G\rangle_\phi,
\label{psie}
\end{eqnarray}
where  $\theta(x)$ is the azimuthal angle of $x$.

The  fluxon part of Lagrangian density  
is given for $a_{xj} = 0$ by 
\begin{eqnarray}
{\cal L}_\phi&=&-\phi^\dagger(\partial_\tau -\mu_\phi)\phi \nonumber  \\
&&  -{V_0\gamma \over 2m}|(\partial_j+i A^{\phi}_{j}
)\phi|^2-H_\phi^{\rm LR}(\phi^\dagger\phi),
\label{Lphi}
\end{eqnarray}
with
\begin{equation}
A^{\phi}_{j}(x) \equiv 2q \int d^2y \frac{\partial}{\partial x_j}
\theta(x-y)(\phi^\dagger\phi(y)-\rho).
\label{CSgauge}
\end{equation}
Here we have included the term 
$H^{\rm LR}_{\phi}$ to ${\cal L}_\phi$, which expresses
a long-range part of the Coulomb repulsion between fluxons. 
We partition the full long-range Coulomb repulsion as follows;
\begin{eqnarray}
H^{\rm LR}&=&\frac{g}{2} \int' d^2xd^2y 
(C^\dagger C(x)-\rho){\ell\over |x-y|}
(C^\dagger C(y)-\rho)\nonumber\\
& \Rightarrow & H^{\rm LR}_\eta +
H^{\rm LR}_\phi, \nonumber\\
H^{\rm LR}_\eta&=&\frac{g_\eta}{2} 
\int' d^2xd^2y (\eta^\dagger \eta(x)-\rho){\ell\over |x-y|}
(\eta^\dagger\eta(y)-\rho), \nonumber\\
H^{\rm LR}_\phi&=& \frac{g_\phi}{2} 
\int' d^2xd^2y
(\phi^\dagger\phi(x)-\rho){\ell\over |x-y|}
(\phi^\dagger\phi(y)-\rho),\nonumber\\
g_\eta + g_\phi &=& g \sim \frac{e^2}{\epsilon \ell},
\label{HLR}
\end{eqnarray}
where the prime indicates that the integration range is
$|x-y| > a$. We note that the recent experiment \cite{geta}
suggests that  the CFs (i.e., chargeons $\eta_x$) have
residual long-range Coulomb repulsion, i.e., $g_\eta \neq 0$.

The above fluxon Lagrangian ${\cal L}_\phi$ has just the 
same form as the Lagrangian of the 
composite bosons (CB) in the CB theory of FQHE\cite{CB}, except
that each fluxon here carries  $2q$ CS flux quanta,
while each CS boson carries $2q+1$ flux quanta.
(We shall discuss the related chargeon-fluxon formalism of  
CB systems
in Sec.\ref{sec:CB}, where $\phi(x)$ carries $2q+1$ flux quanta.)
In Ref.\cite{CB},  the ground state  is given by the Bose 
condensation of CBs and the wave function is calculated explicitly.
Then one may consider $|G\rangle_\phi$ also as the  
Bose-condensed state of $\phi(x)$ as in the CB theory.
By following the derivation in Ref.\cite{CB}, we obtain 
\begin{eqnarray}
&&{}_\phi\langle 0 |\phi(x_1)\cdots \phi(x_N)|G\rangle_\phi  
=\prod_{i<j}|z_i-z_j|^{2q}\ e^{-\frac{1}{4\ell_\phi^2}
\sum_{j=1}^N |z_j|^2},\nonumber\\
&&\ell_\phi\equiv \frac{1}{\sqrt{e B_{\phi}}},\  
B_{\phi} 
= \frac{4\pi q \rho }{e},
\hspace{-0.5cm}
\label{corH}
\end{eqnarray}
where $z_j$'s are the complex coordinates of $N$ fluxons,
$z_j \equiv x_{j1} + i x_{j2}$.
The CS factor $\exp [2iq \int d^2y\theta(x-y)\phi^\dagger_y\phi_y]$ in 
$\tilde{\Psi}_\phi$ 
of (\ref{psie}) produces a phase factor of $|z_i-z_j|^{2q}$,
changing $|z_i-z_j|^{2q}$ in (\ref{corH})
to $(z_i-z_j)^{2q}$ in $\tilde{\Psi}_\phi$ of (\ref{psie}). Thus 
 the fluxon part $\tilde{\Psi}_\phi$ in 
 (\ref{psie}) becomes
\begin{eqnarray}
\tilde{\Psi}_\phi(x_1,\cdots,x_N) 
=\prod_{i<j}(z_i-z_j)^{2q}\ e^{-\frac{1}{4\ell_\phi^2}
\sum |z_j|^2}.
\label{elwave}
\end{eqnarray}
From (\ref{elwave}), one sees that fluxons give rise to the  factor 
$\prod_{i<j}(z_i-z_j)^{2q}$ that describes {\em correlation holes}  
as it is expected.

Let us turn to the chargeon part $\Psi_\eta(x_1,\cdots,x_N)$. 
As explained in Sec.\ref{gtcf}, at $\nu=p/(2pq\pm 1)$, the 
chargeons $\eta_x$ feel the residual field 
\begin{eqnarray}
\Delta B &= & B^{\rm ex} - B_\phi 
= \pm \frac{2\pi \rho}{ep},
\label{deltab2}
\end{eqnarray}
and they fill up the $p$ LL's formed by $\Delta B$, 
giving rise to IQHE.
This observation clearly implies that the chargeons 
are nothing but Jain's CFs.\cite{Jain}
The wave function $\Psi_\eta$ in (\ref{psie}) 
is known for $p=1$ as  the Slater determinant,
\begin{eqnarray}
&&\Psi_\eta(x_1,\cdots,x_N) 
=\prod_{i<j}(z_i-z_j)\ e^{-\sum |z_j|^2/(4\ell_\eta^2)},
\nonumber\\
&&\ell_\eta \equiv \frac{1}{\sqrt{e \Delta B}}.
\label{etawave}
\end{eqnarray}
 Thus (\ref{elwave})
becomes just the Laughlin's wave function for $\nu=1/(2q + 1)$,
\begin{eqnarray}
&&\Psi_e(x_1,\cdots,x_N) 
=\prod_{i<j}(z_i-z_j)^{2q+1}\ e^{-\sum |z_j|^2/(4\ell^2)},\nonumber\\
&& \frac{1}{\ell_\phi^2} + \frac{1}{\ell_\eta^2} = \frac{1}{\ell^2} 
\equiv e B^{\rm ex}.
\label{laughlin}
\end{eqnarray}  
For $p \neq 1$, one
needs the wave function 
$\Psi_\eta$ of IQHE with the filling factor $\nu_\eta= p = 2,3,\cdots$
to obtain the full wave function $\Psi_e$.

At $\nu=1/(2q)\ (p = \infty)$, $\Delta B =0$,
so chargeons behave as quasi-free fermions in vanishing
magnetic field.
Beyond   MFT, fluctuations of $A_{xj}$
meditate interactions among chargeons (and fluxons), and 
may generate non-Fermi-liquid-like behaviors.\cite{kleinert}

\subsection{Low-energy excitations}

Let us study the low-energy excitations in the PFS state
with $\nu=p/(2pq\pm 1)$
at which FQHE is observed experimentally.
There are classified according to the excitaitons in 
the chargeon sector and the excitations in the fluxon sector.
 
We  first consider the excitations in the chargeon sector.
In the leading approximation for PFS in which
the interactions with the gauge field are ignored,
the chargeons move in a 
reduced magnetic field $\Delta B $ of (\ref{deltab})
as explained, and they
fill up the first $p$ LL's formed by $\Delta B$.
So the possible excitations are the  inter-LL excitations
from the $p$-th LL to the $p+1$-th 
LL, hence their energy gap is $\omega_\eta 
\equiv e \Delta B/m_\eta$.
 It is estimated  as 
\begin{eqnarray}
\omega_\eta &=& \frac{e \Delta B}{m_\eta} = \frac{\Delta B}{B^{\rm ex}}
\frac{m}{m_\eta}\omega_B 
\simeq \frac{\nu}{p} \cdot\frac{\nu}{3.2}\cdot 200 {\rm K}
\nonumber\\
&<& T_{\rm PFS} \sim 10 \nu\ {\rm K\ for\ } \frac{\nu}{p} < 0.16,
\end{eqnarray}
where we used the rough estimation, 
$m_\eta/m \sim 3.2/\nu$ (Fig.\ref{METAPHI01}), $T_{\rm PFS} \sim 10 \nu$
(Fig.\ref{T01}) for $g = 0.1 e^2/(\epsilon \ell)$,
and $\omega_B \simeq 200$K. Since $\omega_\eta$ can be smaller 
than $T_{\rm PFS}$, 
these chargeon excitations may be genuine excitations. 

Recently, spin-reversed excitations are observed in the FQH regime
$2/5\ge\nu\ge1/3$.\cite{exp1}
They should be identified with another type of excitations in 
the CF or chargeon sector,
because the CFs may carry spin degrees of freedom.
In fact, if we include the spin degrees of freedom of electrons
into the present chargeon-fluxon formalism, we introduce
the chargeon operator $\eta_{x\sigma}$ which carries the spin index
$\sigma = \uparrow, \downarrow$, corresponding to the electron operator
$C_{x\sigma}$ through (\ref{electronop}) as
\begin{equation}
C_{x\sigma}=\exp [2iq\sum_y\theta_{xy}\phi^\dagger_y\phi_y]
\phi_x\eta_{x\sigma}.
\label{electronopspin}
\end{equation}
Then the observed spin-reversed excitations $|{\rm SR}\rangle$ 
are expected to be Skyrmion-type excitations in the spin space 
of chargeons just like the Skyrmion excitations known in 
the electron system  at $\nu=1$.\cite{skyrmion}
Their excitation energy $E_{\rm SR}$ is estimated as
the sum of Zeeman energy and the Coulomb energy $H^{\rm LR}_\eta$ 
of (\ref{HLR}),
\begin{eqnarray} 
E_{\rm SR} &=& E_{\rm Z} + E_{\uparrow\downarrow},\nonumber\\
E_{\rm Z} &=&  \langle {\rm SR}|H_{\rm Z}|{\rm SR}\rangle,
\nonumber\\
E_{\uparrow\downarrow}&=& \langle {\rm SR}|H^{\rm LR}_\eta
|{\rm SR}\rangle,
\end{eqnarray}
where the Zeeman Hamiltonian is given by 
\begin{eqnarray}
H_{\rm Z} &=& \frac{g\mu_{\rm B} B^{\rm ex}}{2} \sum_x
\Big( C^\dagger_{x\downarrow}C_{x\downarrow} -
C^\dagger_{x\uparrow}C_{x\uparrow}\Big)\nonumber\\
 &=& \frac{g\mu_{\rm B} B^{\rm ex}}{2} \sum_x \phi_x^\dagger \phi_x
\Big( \eta^\dagger_{x\downarrow}\eta_{x\downarrow} -
\eta^\dagger_{x\uparrow}\eta_{x\uparrow}\Big)\nonumber\\
 &\simeq & \frac{g\mu_{\rm B} B^{\rm ex} n }{2} \sum_x
\Big( \eta^\dagger_{x\downarrow}\eta_{x\downarrow} -
\eta^\dagger_{x\uparrow}\eta_{x\uparrow}\Big).
\label{zeeman}
\end{eqnarray}
We have set  $ \phi_x^\dagger \phi_x  \rightarrow
\langle \phi_x^\dagger \phi_x \rangle =n$ in the last line
of (\ref{zeeman}).
From this, we estimate
$E_{\rm Z} \simeq g\mu_{\rm B} B^{\rm ex} n $.
These energies are measured in Ref.\cite{exp1} 
as $E_{\rm Z} \simeq 0.18$mev and 
$E_{\uparrow\downarrow} \simeq 0.6$mev.

Next, let us consider the excitations in the fluxon sector.
Here one may conceive the following two types of excitations.
The first one is described by  small fluctuations  $\varphi_x$ around 
the fluxon condensate (here we consider $T \simeq 0$) as
\begin{equation}
\phi(x)=\sqrt{\rho}+\varphi(x).
\label{smallf}
\end{equation}
To calculate their energy gap, we follow the similar
calculation in CB  theory.\cite{CB,IMO1} 
By substituting (\ref{smallf}) 
into the Lagrangian (\ref{Lphi}), expanding it up to $O(\varphi^2)$,
and making a Bogoliubov transformation to diagonalize it,
one finds that the field $\varphi(x)$ describes excitations
with an energy gap
$\omega_\phi \equiv eB_\phi /m_\phi$.
It is estimated for $g = 0.1 e^2/(\epsilon \ell)$ as
\begin{eqnarray}
\omega_\phi&=& \frac{e B_\phi}{m_\phi} = 
\frac{B_\phi}{B^{\rm ex}}
\frac{m}{m_\phi}\omega_B \nonumber\\
&\simeq& \frac{2pq}{2pq\pm 1} \cdot\frac{\nu}{3} \cdot 200 {\rm K}
\simeq 60\nu\ {\rm K},
\end{eqnarray}
which is always larger than $T_{\rm PFS} \sim 10\nu$. Therefore these
``excitations" are not genuine ones in the PFS states.

The proper fluxon excitations in the PFS states are described 
by the second type of excitations; i.e., vortices.
As in the CB theory of FQHE\cite{CB}, 
they are given by the configurations as
\begin{equation}
\phi(x) \simeq \sqrt{\rho}\ e^{iN\theta(x)},\;\; |x| \gg \ell, 
\label{vortex}
\end{equation}
where $N$ is an integer.
A single vortex carries the electric charge $ceN/(2q)$.
Their kinetic energy vanishes due to
the cancellation of their spatial variaitons with $A^{\rm CS}_{xj}$ 
in the covariant derivative, so an energy gap is 
supplied
from the interaction $H_\phi^{\rm LR}$ 
in (\ref{HLR}). Thus the vortex energy is calculated 
by substituting (\ref{vortex})  into $H_\phi^{\rm LR}$.
The relevant calculation has been already done 
in the CB theory\cite{EV}  for the states of FQHE  at 
$\nu=1/(2q+1)$, 
giving rise to the single-vortex energy $E_V$ for $N=1$ as 
\begin{equation}
E_V \simeq 0.04 {e^2 \over \epsilon \ell} \simeq 6 {\rm K},
\label{EV}
\end{equation}
with $e^2/(\epsilon \ell) \simeq 160$K.
Because this $E_V$ is of the same order as $T_{\rm PFS}$, 
they are in charge of genuine excitations.

The estimation (\ref{EV}) is in agreement 
with the numerical calculation  
by using the Laughlin wave function.\cite{EV,vortex}
However, there are still some discrepancy with the
experimentally observed activation energies in FQH states,
which are smaller than (\ref{EV}) by a factor $2\sim3$. 
From this fact, more careful studies are required to calculate
the energy gap in the fluxon sector in the present formalism.

In the QH states, magneto-roton excitations are known 
as genuine low-energy excitations.
In the present formalism, they are  
identified with the vortex-antivortex pairs in the fluxon
sector just as in the CS theory of CBs. To study these excitaitons,
we start with the effective Lagrangian of vortices obtained by the
duality transformation\cite{zhang}.
Let us consider a system of vortex-antivortex pair with vorticities
$N_i=\pm 1\, (i=1,2)$ and denote their coodinates as 
$(x^1_i, x^2_i)\, (i=1,2)$.
Then the effective Lagrangian is written  after imposing
the LLL condition as
\begin{eqnarray}
L&=& 2E_V -q \ell_\phi^2 \sum _{i=1,2} N_i\epsilon^{\alpha\beta}
\dot{x}^\alpha_i x^\beta_i   \nonumber \\
&&-{1\over 2q}\sum_{i\neq j}\epsilon^{\alpha\beta}\dot{x}^\alpha_i 
{x^\beta_i-x^\beta_j \over |x_i-x_j|^2} 
 -{ e^2 \over 4q^2 |x_1-x_2|}.
\label{Lvor}
\end{eqnarray}
The first term of (\ref{Lvor}) is the self energy of the vortices,
the second term denotes the Lorentz force, the third term comes from
the Aharonov-Bohm phase or fractional statistics of vortices, and 
the fourth term is the Coulomb interaction.
Important results are derived from (\ref{Lvor}).
First the coordinates
$x^\alpha_i$ satisfies the following commutation relations;
\begin{equation}
[x^\alpha_i, x^\beta_j]=2 i q \ell_\phi^2 
N_i\delta_{ij}\epsilon^{\alpha\beta},
\label{CCR}
\end{equation}
which indicate that the sizes of  vortices
are about $\sqrt{2q}\ell_\phi$, and the vortex-pair picture
holds when the distance between vortex and antivortex is lager
than $\sqrt{2q}\ell_\phi$. 
Let us introduce the relative and center-of-mass coordinates,
\begin{eqnarray}
&& x=x^1_1-x^1_2, \; \; y=x^2_1-x^2_2, \nonumber \\
&& X={1\over 2}(x^1_1+x^1_2), \;\; Y={1\over 2}(x^2_1+x^2_2).
\label{xyXY}
\end{eqnarray}
Then from (\ref{CCR}) we have
\begin{equation}
[x,Y]=2iq \ell_\phi^2, \;\; [y,X]=-2iq \ell_\phi^2.
\end{equation}
These relations imply that $(P_X \equiv (2q\ell_\phi^2)^{-1}y,
P_Y \equiv -(2q\ell_\phi^2)^{-1}x )$ 
is the total (center-of-mass) momentum.
From the discussion just below (\ref{CCR}), we  have the conditions,
$P_X, P_Y > (\sqrt{2q}\ell_\phi)^{-1}$.
Let us consider a vortex pair with the total momentum $P_X=p_X$
and $P_Y=0$.
Then the energy of the pair is given from (\ref{Lvor}) as
\begin{eqnarray}
E(y=2qp_X\ell_\phi^2)&=&
2E_V-{e^2 \over (2q)^3 p_X \ell_\phi}.
\label{Epair}
\end{eqnarray}
(One can show that the second and the third terms do not contribute
to $E$.)
Thus  the system has the minimum energy, 
$E_{\rm min} \equiv 2 E_V - e^2(2q)^{-5/2}\ell_{\phi}^{-1}$ at 
$p_X\sim  (\sqrt{2q}\ell_\phi)^{-1}$.
As $y$ or $p_X$ increases to  $\infty$, the vortex pair 
dissociates into an independent vortex and an antivortex,
and its  energy increases continuously up to $2E_V$. 
These independent vortices break the coherent 
condensation of fluxons.

Let us stress that the chargeon excitations with the energy gap
$\omega_\eta$ and the fluxon vortices with the energy gap
$E_V$ (and their magneto-roton combinations) are 
{\it two independent types of excitations}, both
of which are  supported simultaneously
in the PFS states. 
As we shall see in Sec.\ref{transport}, chargeons and 
fluxons contribute to the resistivity tensors, 
$\rho_{xx}$ and $\rho_{xy}$, in different manners.
Therefore, 
by measuring activation energies in two quantities
$\rho_{xx}$ and $\rho_{xy}$,  one may identify 
the above chargeon excitations and 
fluxon vortices independently.
For example, the above mentioned magneto-roton
excitations disappear in the temperature region
$T_{\rm BC}<T<T_{\rm PFS}$, whereas the effective
LLs of CFs, hence the  spin-reversed excitations still exist
and become more active at higher $T$ up to $T_{\rm PFS}$.
Then it is quite interesting to observe by experiments
how the magneto-roton excitations and spin-reversed excitations
change their importance as $T$ increases from very low $T$ 
up to $T_{\rm PFS}$. For  $T_{\rm PFS} < T$, the PFS disappears
and the relevant excitations there should be described by electrons 
themselves. Thus  certain qualitative changes in the
low-energy excitations should be observed at $T = T_{\rm PFS}$.   
Further quantitative study of these excitations and comparison with
experiments are welcome 
to show the appropriateness of the present chargeon-fluxon 
theory of CFs.

Finally,  we comment on the case of $\nu = 1/(2q)$.
In this case,  chargeons move in a vanishing
magnetic field $\Delta B = 0$, so they form
 a Fermi line if
 the interactions with the gauge field are ignored.
Although, the perturbative gauge interactions may 
give rise to a non-Fermi-liquid-like behavior,\cite{kleinert}
 the chargeon excitations may remain  gapless  as 
in a usual Fermi-liquid theory. 
Thus the states are no more incompressive and no FQH effect
will be observed (See next subsection).
In the fluxon sector,  the fluxon vortices with the energy
(\ref{EV}) exist as 
low-energy excitations with a gap 
as in the previous case of $\nu = p/(2pq \pm1)$.

\subsection{EM transport property}
\label{transport}

Let us consider the EM transport properties of the PFS states.
The response functions of electrons is calculated from the EM
effective action $A_{\rm EM}[a_{xj}]$ defined by 
\begin{eqnarray}
\int [dU]\exp(A_{\rm eff}[a_{xj},V_0U_{xj}]) 
= \exp(A_{\rm EM}[a_{xj}]),
\label{SEM2}
\end{eqnarray}
where we have set $\theta_x=0$ as in the previous calculations
and also showed the dependence of $a_{xj}$ explicitly.
In  the PFS states, fluctuations of the dynamical gauge field
$A_{xj}$ are small, so $A_{\rm eff}[a_{xj},V_0U_{xj}]$
can be expanded in powers of $A_{xj}$ up to $O(A^2)$ as
\begin{eqnarray}
&&A_{\rm eff}[a_{xj},V_0U_{xj}]=
-\sum_{x,y,i,j}\Big[(A+cea)_{xi}\Pi^{ij}_{\phi\; xy}(A+cea)_{yj}
\nonumber\\
&&+
(A+(1-c)ea)_{xi}\Pi^{ij}_{\eta\; xy}(A+(1-c)ea)_{yj}\Big],
\label{Seff2}
\end{eqnarray}
where $\Pi^{ij}_{\phi(\eta)}$ is the polarization tensor
of $\phi_x (\eta_x)$.
$A_{\rm EM}[a_{xj}]$ is obtained by making Gaussian  integration   over
 $A_{xj} (\in {\bf R})$  as 
\begin{eqnarray}
A_{\rm EM}[a_{xj}]&=& -e^2 \sum a_{xi}\Pi_{xy}^{ij}a_{yj}, \nonumber\\
\Pi &=& (\Pi_{\phi}^{-1}+\Pi_{\eta}^{-1})^{-1},
\label{resfun}
\end{eqnarray}
where $\Pi$ is nothing but the response function of electrons. 
Then we obtain the simple formula for the resistivity,
\begin{eqnarray}
\rho &=& \rho_\eta+\rho_\phi,\nonumber\\
 \rho &\equiv& \frac{1}{e^{2} \Pi},\ \rho_\eta \equiv
 \frac{1}{e^{2} \Pi_\eta},\ \rho_\phi \equiv
  \frac{1}{e^{2} \Pi_\phi}, 
\label{IL} 
\end{eqnarray}
where $\rho_{(\eta,\phi)}$ are $2 \times 2$
{\em resistivity tensor} of electrons,
chargeons, and fluxons, respectively.
Both chargeons and fluxons contribute additively to $\rho$.
This result obviously reflects the composite nature (\ref{CFoperator})
even though the PFS takes place.

The parameter $c$ in  (\ref{EMcharge})   
expresses arbitrariness to  choose the reference state 
from which the relative EM charges (\ref{EMcharge}) are measured. 
Then it is natural that the formula 
(\ref{IL})  does not depend on $c$.
In high-$T_{\rm c}$ cuprates, there exists exactly the same 
arbitrariness in the EM charges of holons and spinons.\cite{IMO}
There, the formula of $\rho$ that corresponds to (\ref{IL})
is known as the Ioffe-Larkin formula.\cite{IL}

What is the contribution to the electric transports from  fluxons?
Let us consider the FQH state at $\nu=p/(2pq\pm 1)$ first.
In the CB theory of FQHE\cite{CB},  CBs  form 
a Bose condensate and each CB carries
$(2q+1)$-flux quanta and gives rise to $\rho_{xy} =(2q+1)h/e^2$.
Because each fluxon in the present formalism carries $2q$ flux quanta
 and certainly contributes to  
$\rho_{xy}$ in the same manner as those CBs do, 
we have $\rho_{\phi\,xy}=2qh/e^2$.
Likewise,  $\rho_{\phi\,xx}=0$
because of the superfluidity of the fluxon condensation. 
On the other hand, the chargeons fill up the $p$ LL's 
of $\Delta B$ and so
contribute with $\rho_{\eta\,xy} = \pm h/(pe^2)$ and
$\rho_{\eta\,xx} = 0$ as in the IQHE. 
Thus, from (\ref{IL}), we obtain 
\begin{eqnarray}
\rho_{xy}\frac{e^2}{h} &=&2q \pm \frac{1}{p}=\frac{1}{\nu},\ 
\ \rho_{xx} =0,
\label{rhoxy}
\end{eqnarray} 
which are actually observed in the experiments.

We stress that chargeons exhibit IQHE as long as 
$T < T_{\rm PFS}$ whatever fluxons behave as explained at the
end of Sec.\ref{gtcf}.
However, to exhibit FQHE like the Laughlin state (\ref{elwave})
and the resistivity (\ref{rhoxy}), 
fluxons must Bose-condense. Thus FQHE takes place not at
$T < T_{\rm PFS}$ but $T < T_{\rm BC} (< T_{\rm PFS})$.\cite{TBC}

How about the resistivity in the compressible states like $\nu=1/(2q)$?
The fluxons contribute to $\rho_{\phi\,xy}=2qh/e^2$ as before,
whereas the chargeons are effectively in the vanishing magnetic field
and form a Fermi line as explained before.
For such a fermionic system, $\rho_{\eta xx}\neq 0,$ and $\rho_{\eta xy}=0$.
This result shall persist although the gauge field may influences 
the transport properties of the chargeons.
Therefore, the formula (\ref{IL})  becomes
\begin{equation}
\rho_{xy}\frac{e^2}{h} =2q=\frac{1}{\nu},\ \ \rho_{xx} \neq 0,
\end{equation} 
 as is also observed experimantally.

Finally, let us point out the possible change in the behavior of 
$\rho_{xx}(T)$ across $T =  T_{\rm PFS}$, which reflects the change 
of quasiparticles from electrons  for
$T_{\rm PFS} < T$ 
to (uncondensed) fluxons (and chargeons and perturbative gauge bosons)
at $T < T_{\rm PFS}$.
For  $T_{\rm BC} < T < T_{\rm PFS}$,
chargeons still give rise to
$\rho_{\eta\; xx} \simeq 0 $ 
(up to $T_{\rm PFS}$) due to their IQHE 
($\rho_{\eta\; xx} \propto \exp(-\beta\omega_\eta)$ by the
small LL mixing effects).
However, vortices of fluxons are activated there and 
the Bose condensation disappears. Thus  the
 fluxons give rise to a nonvanishing 
contribution $\rho_{\phi\; xx} \neq 0$, and the FQH states disappear
as we mentioned above.
If this change is observed in experiment, 
it may support our present theory of PFS. 
We recall that a similar change in high-$T_{\rm c}$ cuprates
is observed, where
anomalous behaviors  of various physical quantities start 
to appear at $T=T_{\rm CSS}$, 
below which the anomalous metallic phase is realized.
For example, the dc resistivity exhibits a linear-$T$ behavior
below $T_{\rm CSS}$, which is consistent with the perturbative 
calculations
using a system of bosons interacting with a massless gauge field.

Let us discuss the above observation rather in detail.
For $T_{\rm PFS} < T$, the electrons in $B_{\rm ex}$ 
interact themselves and with impurites, producing  certain
$\sigma_{e\; xx}$ and $\sigma_{e\; xy}$.
For $T$ just below $T_{\rm PFS}$, the fluxon degrees of freedom 
is well described not by $\phi_x$ but by the following 
$\tilde{\phi}_x$ because of the {\em noncondensation}
of the field $\phi_x$;
\begin{eqnarray}
\tilde{\phi}_x \equiv \exp(-2iqn\sum_y \theta_{xy}\phi_y^\dagger 
\phi_y)\phi_x,
\end{eqnarray}
with which the interaction term in $J_{xj}$  of 
(\ref{J}) is expressed as 
\begin{eqnarray}
J_{xj}&\equiv& {1 \over 2ma^2}\Big(\gamma\tilde{\phi}_{x+j}
\exp(-2iqn\nabla_j \sum_y \theta_{xy})e^{ieca_{xj}}
\tilde{\phi}_x^\dagger\nonumber \\
&&+\cdots \Big).
\label{J2}
\end{eqnarray} 
This shows that the new bosons $\tilde{\phi}_x$ move
under $B_{\phi}$. Such a {\it bosonic} system may have smaller 
$\sigma_{xx}, \sigma_{xy}$ than those of the electron system (above 
$T_{\rm PFS}$) due to the larger bosonic density of states
and {\em the interaction with the dynamical gauge field} $U_{xj}$, i.e., 
$\sigma_{\phi\; xx}(T) = f(T) \sigma_{e\; xx}(T),\
\sigma_{\phi\; xy}(T) = g(T) \sigma_{e\; xy}(T)$, 
$f(T) <  1, g(T) < 1$.
In the gauge theory for the t-J model in the slave-boson
picture, the conductivity of the bosonic holon and the fermionic
spinon, which are interacting with the dynamical gauge field, 
is calculated\cite{tJ}.
At low $T$, the holon gives main contribution to the resistivity as
$\sigma \propto T^{-1}$ which should be compared with the ordinary
$T^{-2}$ behavior at low $T$.

\begin{figure} 
  \begin{picture}(180,130) 
    \put(40,0){\epsfxsize 130pt \epsfbox{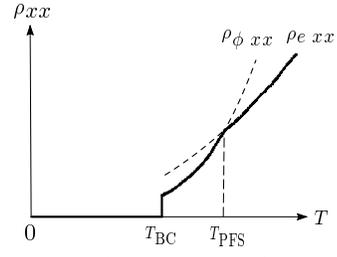}} 
  \end{picture} 
  \caption{Expected $T$-dependence of $\rho_{xx}$ at a fixed $\nu$. 
  \newline
  For $0 < T < T_{\rm BC}$, $\rho_{xx}=0$ because  
  $\rho_{\eta\; xx}=0$ and  $\rho_{\phi\; xx}=0$.
  For $  T_{\rm BC}< T < T_{\rm PFS}$,
  $\rho_{\eta\; xx} = 0$ (ignoring small LL mixings) but
$\rho_{\phi\; xx} \neq 0$,
  since Bose condensation at $T < T_{\rm BC}$ disappears here.
 For $T_{\rm PFS} < T$ the quasiparticles changes to electrons,
which may generate the different $T$-behavior.
If we assume $\rho_{xx}$ is continuous at $T = T_{\rm PFS}$,
we expect a discontinuity in  $d\rho_{xx}(T)/dT$ at $T = T_{\rm PFS}$.
The curve of $\rho_{\phi\; xx}$ is drawn for  
$r(T) = (T/T_{\rm PFS})^{m-2n}$ by assuming  $m > 2n$. 
  }
\label{rhoxx}
\end{figure}

To study the behavior of $\rho_{xx}(T)$ near $T = T_{\rm PFS}$, 
let us start with the formula of $\rho=\sigma^{-1}$;
\begin{eqnarray}
\left(
\begin{array}{cc}
\rho_{xx}  & \rho_{xy}\\
\rho_{yx} & \rho_{yy}
\end{array}
\right) =
 \frac{1}{\sigma_{xx}^2+\sigma_{xy}^2}
\left(
\begin{array}{cc}
\sigma_{xx}  & -\sigma_{xy}\\
\sigma_{xy} & \sigma_{xx}
\end{array}
\right),
\end{eqnarray} 
where we set $\sigma_{yx} = - \sigma_{xy}, \sigma_{yy} = \sigma_{xx}$.
It may be simplified by setting $\sigma_{xx} << \sigma_{xy}$ as 
\begin{eqnarray}
\rho_{xx} \simeq \frac{\sigma_{xx}}{\sigma_{xy}^2}.
\end{eqnarray} 
Thus the ratio of $\rho_{\phi\; xx}$ and $\rho_{e\; xx}$
is expressed as
\begin{eqnarray} 
r(T) \equiv \frac{\rho_{\phi\; xx}(T)}{\rho_{e\; xx}(T)} 
= \frac{f(T)}{g^2(T)}.
\end{eqnarray} 
At $T=T_{\rm PFS}$ we assume $\rho_{xx}$ is continuous;
$r(T_{\rm PFS})  = 1$.
For example, let us set the $T$-dependences  $f(T) \propto T^m,
g(T) \propto T^n$. Then we have
\begin{eqnarray} 
r(T) \simeq   \left(\frac{T}{T_{\rm PFS}}\right)^{m-2n},
\end{eqnarray} 
which implies that the derivative $d\rho_{xx}(T)/dT$
is discontinuous at $T = T_{\rm PFS}$.  According to
 $m > 2n$ or $m < 2n$,
$\rho_{xx}$ gets reduction $r(T) < 1$ or enhancement
$r(T) > 1$ at $T < T_{\rm PFS}$.
On the other hand,
at $T$ slightly higher than $T_{\rm BC}$, we expect 
$\rho_{\phi\,xy}\simeq 2qh/e^2$
and $\rho_{\phi\,xx} \simeq 0$.
In Fig.\ref{rhoxx} the expected $T$-dependence of $\rho_{xx}$ is
illustrated.

\section{Discussion}
\setcounter{equation}{0}

\subsection{Conclusion}

\begin{figure} 
  \begin{picture}(200,160) 
    \put(0,0){\epsfxsize 220pt \epsfbox{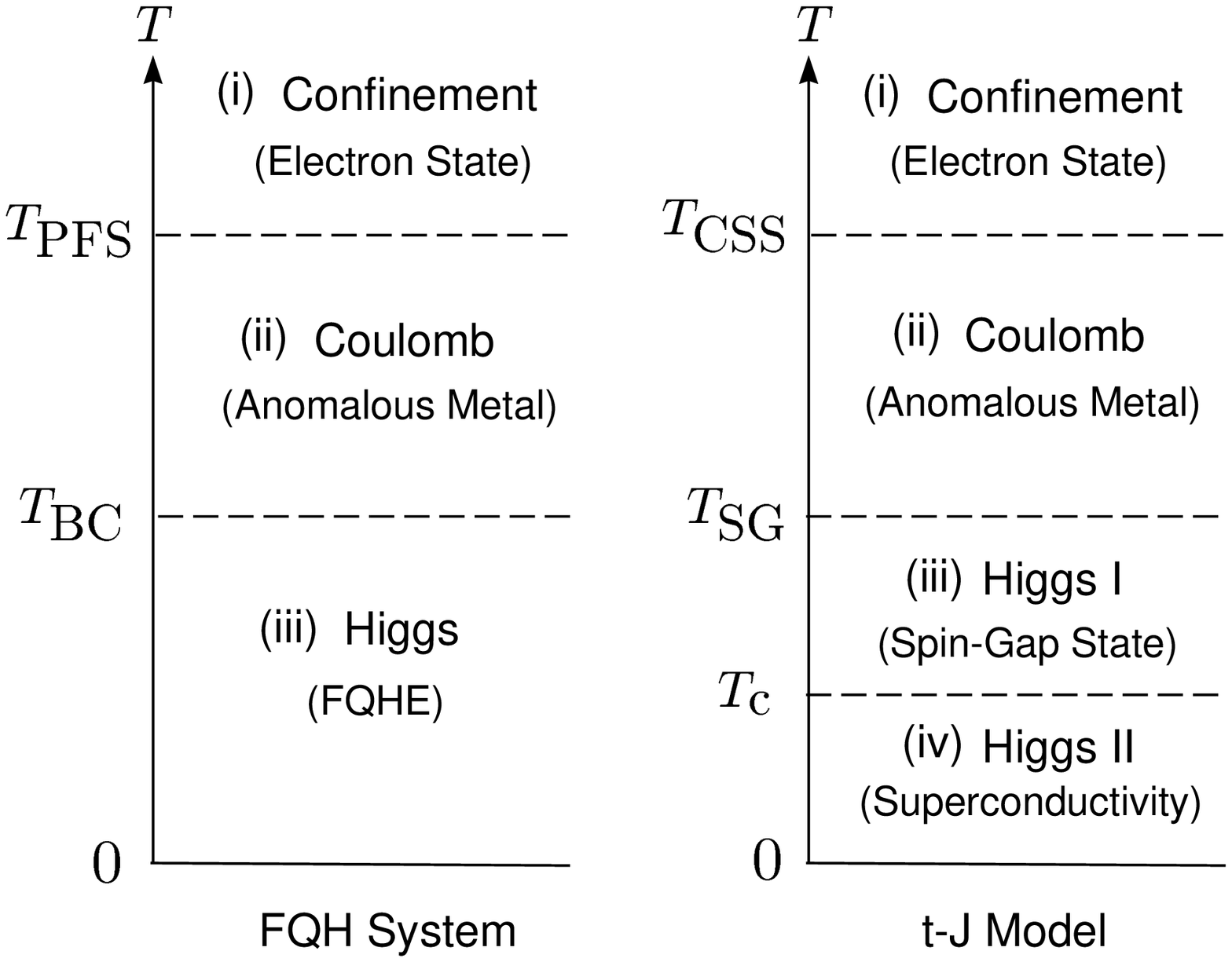}} 
  \end{picture} 
  \caption{Phase structure of the gauge dynamics along the
  temperature axis $T$ for the 
  FQH system studied in the present paper (left) 
  and the t-J model of high-$T_{\rm c}$ cuprates\cite{CSS}
  (right). In the t-J model, (i) holons and spinons
  are confined in electrons above  $T_{\rm CSS}$;
  (ii) CSS takes place below
  $T_{\rm CSS}$, where holons and spinons are deconfined, exhibitting
  anomalous-metallic behaviors;.
  (iii) the spin-gap develops below $T_{\rm SG}$;
  (iv) holons Bose-condense below $T_{\rm c}$ and
  induce the superconductivity (for some values of holon density).
  In FQH system, (i) chargeons and fluxons 
  are confined in electrons above $T_{\rm PFS}$;
  (ii) chargeons and holons are deconfined below $T_{\rm PFS}$ and
  some anomalous behaviors are expected; (iii) fluxons Bose condense
  and FQHE takes place.  There is an almost complete correspondence 
  between the gauge dynamics of these two models.
  }
\label{phase}
\end{figure}

In this paper we studied the low-energy quasiexcitations of the
fermionic CS gauge theory for the CF.
We show that phenomenon which we call PFS is essential for the
CF to appear as a quasiparticle and the PFS takes place at low 
$T<T_{\rm PFS}$.
The PFS can be understood as a deconfinement transition of the
dynamical gauge field.
As a result of the PFS, an electron or a CS fermion splinters off
a chargeon and a fluxon. Below $T_{\rm BC}$, fluxons
Bose condense,  cancelling the external magnetic field partly
and producing correlation holes in the desired form.
The chargeons move in a reduced manetic field
$\Delta B$ to form an incompressible fluid 
as  the Jain's CFs.
The system below $T_{\rm BC}$ exhibit FQHE.
We estimated the transition temperature $T_{\rm PFS}$
as well as the CF and fluxon masses.
From the local gauge invariance with the dynamical gauge field,
chargeons and fluxons contribute to the resistivity tensor of electrons 
additively. 
This formula describes well the observed results of resistivity.
For the important problem in future, we consider the calculation of
$T_{\rm BC}$. It is not only important for comparison with experiments
but also challenging theoretically.
To compare with the experiments, we have calculated 
$T_{\rm PFS}$, $m_\eta$, $m_\phi$. Also we conjectured $\rho_{xx}$
should have an extra damping factor below $T_{\rm PFS}$. 
Finally, we stress that PFS is closely related to CSS in  
high-$T_{\rm c}$ cuprates.
Actually, theoretical techniques are common and 
both phenomena are understood as separations of 
the degrees of freedom describing electrons. 
In Fig.\ref{phase} we illustrate the phase strucutres
of the gauge dynamics for the present FQH sysetem and the 
t-J model of high-$T_{\rm c}$ cupper oxides.\cite{CSS,IMS,IMO}
We note that a deconfinement phase may be further classified into
two independent phases; (I) Coulomb phase in which
the gauge bosons are massless and (II) Higgs phase
in which the gauge bosons acquire a finite mass.
The two phase strucures in Fig.\ref{phase} look 
almost similar each other, except that 
the present FQH system has one Higgs phase that is generated
by the Bose condensation of fluxons via the well-known
Anderson-Higgs mechanism, while 
the t-J model has two Higgs phases 
reflecting the two different mechanisms\cite{CSS,IMS,IMO};
(I) Higgs I induced by the spin-gap order parameter  and (II)
Higgs II induced by the Bose condensation of holons.

\subsection{PFS in composite bosons }
\label{sec:CB}

Let us comment on the CB approach to the FQHE in terms of
the ``chargeon" and ``fluxon".
First we  intorduce the CS boson operator $\psi^{\rm B}_x$ 
from the electron operator $C_x$ by the
following CS transformation;
\begin{equation}
C_x=\exp[(2q+1)i \sum_y \theta_{xy} 
{\psi_y^{\rm B}}^\dagger\psi^{\rm B}_y]\psi^{\rm B}_x.
\label{CSBtrans}
\end{equation}
Each CS boson is viewed as a composite of an electron and
$(2q+1)$ quanta of CS fluxes. Then the ``chargeon" operator
$\eta^{\rm B}_x$ and the ``fluxon" operator $\phi^{\rm B}_x$ are
intoroduced as constituents of a CS boson
just like the case of 
CS fermion in (\ref{CFoperator}) as follows;
\begin{equation}
\psi^{\rm B}_x=\phi^{\rm B}_x\eta^{\rm B}_x,
\label{CB}
\end{equation}
where both $\phi^{\rm B}_x$ and $\eta^{\rm B}_x$ are canonical bosons. 
Each fluxon
$\phi^{\rm B}_x$ carries an {\em odd} number $(2q+1)$ of flux quanta.
One may follow the analyses of Sec.II and Sec.III and
obtain a critical temperature $T^{\rm B}_{\rm PFS}$
below which the gauge dynamics is realized in a deconfinement
phase and ther PFS takes place.

In the FQH states, both chargeons and fluxons Bose condense
and  fluxons create the correlation holes.
In this sense, the fluxon field corresponds to the longitudinal
gauge field in the SM formalism of the CB.
However, the PFS is essential for the chargeon and fluxon to behave as
weakly interacting quasiexcitations and to satisfy the usual canonical
commutation relations.

\subsection{Comment on the theory of Shanker and Murthy}

As we mentioned in the introduction, recently there appeared
some papers for the low-energy quasiexcitations in the FQH state
by using the CS theory.\cite{SM}
Let us briefly discuss the SM's approach among them, in particular, 
the treatment of the CS constraint and field operators which
they introduced in order to describe quasiparticles.

Lagrangian of the CS theory is given as follows,
\begin{eqnarray}
L&=& i\bar{\psi}\partial_0 \psi+a_0
\Big({\vec{\nabla} \times \vec{a} \over 2\pi l}
-\bar{\psi}\psi\Big)  \nonumber   \\
&&-{1\over 2m}|(-i\vec{\nabla}+e\vec{A}+\vec{a})\psi|^2,
\label{SML}
\end{eqnarray}
where $\psi$ is the CS particle field (a boson or fermion), $\vec{A}$ is the
external magnetic field and $a_0$ is the Lagrange multiplier
for the CS constraint,
\begin{equation}
{\vec{\nabla} \times \vec{a} \over 2\pi l}
-\bar{\psi}\psi=0.
\label{SMCSC}
\end{equation}
Then SM introduced {\em a composite particle operator} $\psi_{CP}$;
\begin{eqnarray}
\psi(x,y,t)&=&\exp [i\Theta(x,y,t)]\psi_{CP}(x,y,t), \label{CPO} \\
\Theta(x,y,t)&=&-\int^t_{-\infty} a_0(x,y,t')dt'.
\label{Theta}
\end{eqnarray}
The Lagrangian (\ref{SML}) is rewritten in terms of $\psi_{CP}$,
\begin{eqnarray}
L&=& i\bar{\psi}_{CP}\partial_0 \psi_{CP}+a_0
{\vec{\nabla} \times \vec{a} \over 2\pi l}  \nonumber   \\
&&-{1\over 2m}|(-i\vec{\nabla}+e\vec{A}+\vec{a}+2\pi l\vec{P})\psi_{CP}|^2,
\label{LCP}
\end{eqnarray}
where 
\begin{equation}
\vec{P}(x,y,t)={1\over 2\pi l}\vec{\nabla} \Theta(x,y,t).
\label{P}
\end{equation}
From (\ref{Theta}) and (\ref{P}),
\begin{equation}
\vec{\nabla}^{-1}\cdot\partial_0 \vec{P}(x,y,t)=-{1\over 2\pi l}a_0(x,y,t).
\label{Pa0}
\end{equation}
By substituting (\ref{Pa0}) into the second term of (\ref{LCP}),
it is easily seen that the CS gauge field $\vec{a}$ and $\vec{P}$
become canonical conjugate variables with each other.
More precisely let us define the following variables
after going into the momentum space $\vec{q}$,
\begin{equation}
P=-i {\vec{q} \over q}\cdot \vec{P}, \;\; 
\delta\vec{a}=e\vec{A}+\vec{a},
\;\; \delta a=i{\vec{q}\over q}\times \delta\vec{a}.
\label{newv}
\end{equation}
Then $[\delta a(\vec{q}_1),P(\vec{q}_2)]=(2\pi)^2 
\delta^2(\vec{q}_1+\vec{q}_2)$.

The local CS constraint on ``physical states" appears
from the invariance under time-independent gauge transformations; 
$\psi_{CP} \rightarrow e^{i\Lambda}\psi_{CP}$ and 
$\vec{P} \rightarrow \vec{P}-{1\over 2\pi l}\vec{\nabla} \Lambda$,
\begin{equation}
\Big({\vec{\nabla} \times \delta\vec{a} \over 2\pi l}
-:{\psi}^\dagger_{CP}\psi_{CP}:\Big)|\rm phys\rangle=0,
\label{SMCSC2}
\end{equation}
where $:{\psi}^\dagger_{CP}\psi_{CP}:={\psi}^\dagger_{CP}\psi_{CP}-\rho$
and SM used (\ref{SMCSC2}) for deriving the Laughlin's wave
function, etc.\cite{SM}(see later discussion).

In the ``new" Lagrangian (\ref{LCP}), SM treated $\psi_{CP}$ and
$\vec{P}$ as {\em independent commuting dynamical variables} instead of
the ``original" ones $\psi$ and $a_0$.
But this treatment is {\em not} legitimate by the following
reason;
From (\ref{CPO}) and the fact that $\vec{P}$ and $\vec{a}$
are conjugate operators, the new variable $\psi_{CP}$ must satisfy
{\em a nontrivial nonlocal commutation relation} with the CS gauge 
field $\vec{a}$ or $\delta a$ defined by (\ref{newv}).
Therefore $\psi_{CP}$ cannot be treated as an independent variable
that commute with other operators.
In other words, SM have {\em changed} the system from the original one.

From (\ref{CPO}), it is obvious that the above local gauge symmetry
comes from the invariance of the original field $\psi$ under
a simultanoeous phase rotation of $\psi_{CP}$ and $\exp(i\Theta)$.
Therefore, this gauge symmetry is close to the one in the 
chargeon-fluxon approach in this paper.
As we discussed in this paper, the most important point in the gauge theory 
is how the
above local gauge symmetry and the constraint (\ref{SMCSC2}) 
are realized in the ground state and low-energy excitations.\cite{IMO1}
This dynamical problem is essential and must be clarified.
For example in QED, the Gauss' law constrant, which is similar to
(\ref{SMCSC2}), is {\em not} satisfied by the low-energy excitations,
i.e., the electron and photon.
Similarly, in the fermionic CS gauge theory, the multiplier
$a_0$ often acquires a ``mass term" from the radiative corrections
of the fermion $\psi$ and the CS constraint becomes 
irrelevant at low energies.

Let us consider the case of CS bosons and examine the SM's derivation of the
Laughlin's wave function for $\nu=1/3$.
By ignoring the interaction terms of the ``gauge field" $(a,P)$ and 
the composite boson $\psi_{CP}$ as SM did, Hamiltonian is given as follows
from (\ref{LCP}),
\begin{equation}
H_0={1\over 2m}|\vec{\nabla}\psi_{CP}|^2+{\rho \over 2m}(\delta
a^2+(6\pi)^2P^2).
\label{HCP}
\end{equation}
If we {\em assume} that $\psi_{CP}$ and $(\delta a,P)$ are independent
variables as SM did,
wave functional of the lowest-energy state of the Hamiltonian (\ref{HCP}),
$\Psi[a,\psi_{CP}]$, is readily obtained.
The boson field $\psi_{CP}$ Bose condenses and the system of $(a,P)$ is
just a harmonic oscillator (here we ignore quantum fluctuations of
$\psi_{CP}$ as SM did),
\begin{equation}
\Psi[a,\psi_{CP}]=\prod_x\delta(\psi_{CP}(x)-\sqrt{\rho})
\cdot \prod_{\vec{q}}e^{-\delta a^2(\vec{q})}.
\label{waveF}
\end{equation}
However it is obvious that the wave functional (\ref{waveF}) 
does {\em not} satisfy the constraint (\ref{SMCSC2}).
The Bose condensation of $\psi_{CP}$ breaks the gauge invariance.
Moreover, the Hamiltonian $H_0$ in (\ref{HCP}) itself does {\em not}
respect the local gauge symmetry generated by the operator
$\Big({\vec{\nabla} \times \delta\vec{a} \over 2\pi l}
-:{\psi}^\dagger_{CP}\psi_{CP}:\Big)$.
SM simply ignored the $\psi_{CP}$ part of the wave functional and simply
put $\Psi[a,\psi_{CP}]=1\cdot \prod_{\vec{q}}e^{-\delta a^2(\vec{q})}$
where $1$ stands for ``wave functional" of the Bose condenesd state.
This manipulation is essential for SM's derivation of the Laughlin's
wave function.
Obviously they confused the field operator in the second quantization 
and the wave function.\cite{boseC}
From the above discussion, it is obvious that SM's treatment of the
CS constraint is {\em not} satisfactory at all.

Contrary to the SM approach, new dynamical degrees of freedom 
for the CS fluxes is introduced in the present formalism, i.e., the
fluxon field.
The fluxon field can be quanatized as canonical bosons and it
 commutes with other
fields like the chargeon and the gauge field.
As the phase factor $\exp(i\Theta)$ in the SM approach, the fluxon
produces the correlation holes and contains the reminiscence 
of the inter-LL excitations.

As we stressed in the present paper, PFS is essential to assure
the validity of Jain's CF theory (such as the stability of CFs),
and to identify the conditions with which PFS is possible
is a purely dynamical problem. In this context, we recall there is
a similar dynamical problem of gauge theory that is studied as
``hidden gauge symmetry"\cite{HGS} in a context of 
composite partcle theory in high-energy 
particle physics.\\

{\bf Acknowledgement}\\

We dedicate this paper to the memory of Professor Bunji Sakita.
He has influenced us through his attitude to research and his character.


\newpage

\begin{references} 
 
\bibitem[*]{ichinose} Electronic address: 
ikuo@ks.kyy.nitech.ac.jp 
\bibitem[\ddagger]{matsui} Electronic address: 
matsui@phys.kindai.ac.jp 


\bibitem{Jain}J.K.Jain,
Phys.Rev.Lett.63, 199(1989).

\bibitem{LF}A.Lopez and E.Fradkin,
Phys.Rev.B44, 5246(1991).

\bibitem{HLR}B.I.Halperin, P.A.Lee, and N.Read,  \\
Phys.Rev.B47, 7312(1993).

\bibitem{SM}R.Shankar and G.Murthy,
Phys.Rev.Lett.79, 4437(1997); G.Murthy and R.Shankar:
{\em Composite Fermions}, ed. O.Heinonen (World Scientific,
Singapore, 1998).

\bibitem{PFS1}I.Ichinose and T.Matsui,
Nucl.Phys.B468,487(1996).

\bibitem{PFS2}I.Ichinose and T.Matsui,
Nucl.Phys.B483, 681(1997).

\bibitem{CSS}I.Ichinose and T.Matsui,
Nucl.Phys.B394, 281 (1993); \\
Phys.Rev.B51, 11860 (1995).

\bibitem{IMS}
I.Ichinose, T.Matsui  and  K.Sakakibara,    
J. Phys. Soc. Jpn. 67, 543 (1998).  

\bibitem{IMO}I.Ichinose, T.Matsui, and M.Onoda, Phys.Rev.B64, \\
104516(2001); 
See also I.Ichinose and T.Matsui,\\
Phys.Rev.Lett.86, 942(2001).

\bibitem{PFS3}I.Ichinose and T.Matsui, Jour. Phys. Soc. Jpn.
71, 1828 (2002).


\bibitem{wilson}
K.Wilson, Phys.Rev.D10, 2445(1974).

\bibitem{fradkin}
E.Fradkin,Phys.Rev.Lett.63,322(1989);
T.Matsui, K.Sakakibara, and H.Takano,
Phys.Lett..A194, 413(1994).

\bibitem{sakita}
B.Sakita, {\em Quantum Theory of Many Variable Systems and
Fields}, (World Scientific, Singapore, 1985).

\bibitem{Nayak}C.Nayak,
Phys.Rev.Lett.85, 178(2000).


\bibitem{QCD}Y.Iwasaki,K.Kanaya, S.Sakai, and T.Yoshie, \\
Phys.Rev.Lett.69, 21(1992).

\bibitem{HCboson}The fluxons can be consistently quantized 
as  hard-core bosons instead of ordinary bosons.\cite{PFS1,PFS2}
In this case, the condition (\ref{gamma}) is changed to
$(2ma^2)^{-1}(\gamma^2+\gamma^{-2}) \simeq e^2/(\epsilon \ell)$.

\bibitem{Definition}Here we once again emphasize that   ambiguity of
the parameters $\gamma, g_1, g_2$ corresponds to the freedom of 
in definitions of chargeons and fluxons.
We have determined these parameters by the condition that the 
quasiparticles must be free as much as possible.

\bibitem{CB}
S.C.Zhang, T.H.Hansson, and S.A.Kivelson, 
Phys. Rev. Lett. 62, 82(1989);
N.Read, Phys.Rev.Lett.62, 86(1989);
Z.F.Ezawa, M.Hotta, and A.Iwazaki, Phys.Rev.B46, 7765 (1992).


\bibitem{CD}A.M.Polyakov, Phys.Lett.B72, 477(1978); 
L.Susskind, Phys.Rev.D20, 2610(1979).
See also Sec.IIB of the first reference of Ref.\cite{IMO}.

\bibitem{hopping}
Strictly speaking, to obtain the  full Ginzburg-Landau theory
of $U_{xj}$,
one needs to calculate the full 
propagators of $\eta_x$ and $\phi_x$ that include 
the effects of all the possible hoppings with arbitrary
numbers of steps. The present hopping expansion
keeps just the leading terms with the shortest pathes. 
The effect of long contours may renormalize the quantitative
results such as $T_{\rm PFS}$ obtained in the hopping
expansion, although the qualitative results are unchanged.   

\bibitem{correction}
We note that, in the case that the fluxons are hard-core 
bosons \cite{PFS2}
  instead of canonical bosons, $D_2$ in $A_2$ of (\ref{S2}) is to be
 replaced by $D'_2 \equiv n(1-n) 
 (\gamma^2 + \gamma^{-2})/(4m^2a^4)$.
In Ref.\cite{PFS3}, the expression of $S_2(\equiv -A_2)$ 
for the canonical fluxons, Eq.(13), contained $D'_2$ incorrectly, 
which should read $D_2$. Therefore, the numerical results of
$T_{\rm PFS}, T_{V_0}, m_\eta, m_\phi$ 
for $\nu = 1/2$ given in Ref.\cite{PFS3} should be replaced
by the correct values of   (\ref{TPFS2},\ref{TV01/2},\ref{CFmass}) 
calculated in Sect.\ref{numericalresults}.
However, the differences are almost negligible.

\bibitem{PFSexp} R.L.Willett, R.R.Ruel, M.A.Paalanen, K.W.West, 
and L.N.Pfeiffer, Phys.Rev.B47, 7344 (1993).


\bibitem{kleinert}H.Kleinert, F.S.Nogueira, and A.Sudbo,
Phys.Rev.Lett.88, 232001(2002). See also
C.Nayak anf F.Wilczek, Nucl.Phys.B417, 359(1994); B430, 534(1994);
M.Onoda, I.Ichinose, and T.Matsui, Nucl.Phys.B446, 353(1995).

\bibitem{antonov}D.Antonov,
Phys.Lett.B535, 236(2002) and references cited therein.

\bibitem{geta}
K.Kumada et.al. Phys.Rev.Lett.89, 116802(2002).

\bibitem{IMO1}I.Ichinose, T.Matsui and M.Onoda, \\
Phys.Rev.B52,10547(1995).

\bibitem{exp1}I.Dujovne, A.Pinczuk, M.Kang, B.S.Dennis, L.N.Pfeiffer,
and K.W.West,
Phys.Rev.Lett.90, 036803(2003).

\bibitem{skyrmion}
See, e.g., M.Shayegan, ``Les House Lecture Note" pp.3 (1998)
ed. by A.Comdet et al.

\bibitem{EV}See, e.g., the third reference of Ref.\cite{CB}.

\bibitem{vortex}
R.Morf and B.I.Halperin, Phys.Rev.B 33, 2221(1986);
B.I.Halperin, Surf.Sci.170, 115(1986).

\bibitem{zhang}S.C.Zhang,
Int.J.Mod.Phys.B6, 25(1992); \\
See also
I.Ichinose and A.Sekiguchi,
Nucl.Phys.B493[FS], 683(1997).

\bibitem{IL}L.B.Ioffe and A.I.Larkin,
Phys.Rev.B39, 8988(1989).


\bibitem{TBC}
In this paper, we have assumed $T_{\rm BC} < T_{\rm PFS}$, which
is logically necessary for the PFS theory.
One may think that the BC occurs only at $T=0$ in the 
pure $2D$ systems.
However if the boson couples with a CS gauge field and there
is a replusion between bosons as in the present fluxon system, 
all excitations have energy gaps.
Then it is possible that $T_{\rm BC}\neq 0$.
Actually, as far as we know, there is no reliable estimation of $T_{\rm BC}$
for such systems.
Both $T_{\rm PFS}$ and $T_{\rm BC}$ are controlled by the repulsive 
interactions between electrons, and we need a convincing 
calculation of $T_{\rm BC}$.

\bibitem{tJ}P.A.Lee and N.Nagaosa,
Phys.Rev.B 46, 5621 (1992); 
M.Onoda, I.Ichinose, and T.Matsui,
Phys.Rev.B 54, 13674 (1995).

\bibitem{boseC}In the first-quantization representation,
one-body wave functions of the Hamiltonian $m^{-1}
|\vec{\nabla}\psi_{CP}|^2$ are obtained as 
$\psi_{CP}=\exp(i\vec{q}\cdot \vec{x})$.
Then the wave function of the Bose-condensed state in many-body system,
$\Phi(\vec{x}_i)$, is simply given as $\Phi(\vec{x}_i)=1$.
However in this representation, the most important aspect of the
Bose condensation, i.e., the spontaneous breaking of the gauge symmetry,
cannot be clearly seen.


\bibitem{HGS}M.Bando, T.Kugo and K.Yamawaki,
Phys.Rep.164, 217(1988).

\end{references}
\end{document}